# Fast Localized Urban Microclimate Emulator: Wind and Temperatures in Unseen Urban Morphologies

Shaoxiang Qin[1,2], Theodore Potsis[1], Dongxue Zhan[1], Xue Liu[2], Ted Stathopoulos[1], and Liangzhu Leon Wang[1,*]

[1]Concordia University, Centre for Zero Energy Building Studies, Department of Building, Civil and Environmental Engineering, Montreal H3G 1M8, Canada

[2]McGill University, School of Computer Science, Montreal, H3A 0E9, Canada

## Highlights

- FLUME predicts 3D urban wind and temperature using only 23 training blocks
- Visibility-based geometry features capture adaptive spatial scales
- Unseen-case MAEs are 0.2 m/s for streamwise velocity and 0.19°C for temperature
- Ensembles and conformal prediction yield calibrated prediction intervals
- Datasets and a live interactive demo are publicly available at flume.cityffd.ai

## Abstract

Urban microclimate, encompassing wind and temperature fields shaped by building geometry, strongly affects building energy use, pedestrian wind comfort, pollutant dispersion, the urban heat island, and public health. Conventional Computational Fluid Dynamics (CFD) is computationally prohibitive for rapid assessments, while many deep learning approaches require extensive training data and struggle with generalization to unseen configurations. In this work, we present the Fast Localized Urban Microclimate Emulator (FLUME), a rapid, data-efficient, and geometry-aware framework based on a Fourier neural operator for predicting three-dimensional wind and temperature fields. FLUME assumes that, under prescribed inflow and thermal forcing, much of the spatial variation of these mean fields is controlled by surrounding building geometry. The framework introduces a Multi-Directional Distance Feature (MDDF), which captures geometry over adaptive, visibility-defined spatial scales. By decomposing simulations of only 23 urban blocks into smaller 3D patches enriched with precomputed MDDF information, FLUME enables learning from limited CFD data. The model achieves mean absolute errors of 0.2 m/s for streamwise velocity and 0.19°C for temperature on unseen urban configurations. A deep ensemble provides a practical proxy for epistemic uncertainty, and conformal prediction transforms these estimates into calibrated prediction intervals. Ensemble spread is positively related to local prediction error, with greater uncertainty in roof-level shear layers, building wakes, and vertical velocity near the ground. These results indicate FLUME's potential for rapid scenario assessment in urban design and urban climate studies.

## 1. Introduction

The interaction between the Atmospheric Boundary Layer (ABL) and the urban environment, shaped by building geometry, density, and spacing, governs phenomena such as pedestrian wind and thermal comfort, thermal stress, pollutant dispersion, and the urban heat island (UHI) effect (Toparlar et al., 2017; Yang et al., 2023). Understanding and predicting these phenomena accurately is therefore a prerequisite for evidence-based urban planning, sustainable building design, and resilient public health policy. The thermal and aerodynamic conditions in cities are tightly connected: outdoor wind flow controls the ventilation of building facades and street canyons, influences heating and cooling loads (Brozovsky et al., 2022; Hong et al., 2021), and modulates the dispersion of heat and pollutants (Tominaga and Stathopoulos, 2013; Boikos et al., 2025). Thus, a fast and reliable prediction of urban microclimate fields has become an increasingly important research target.

Computational Fluid Dynamics (CFD) has long been the reference tool for urban microclimate simulation. Its use for environmental applications has rapidly expanded during the last three decades, as documented in several review studies (Blocken, 2014; Potsis et al., 2023). RANS and LES methods have been shown to resolve building-induced flow patterns with good agreement against wind tunnel and field measurements (Tominaga et al., 2008a; Antoniou et al., 2019; Franke et al., 2011; Potsis et al., 2024). Potsis et al. (2024) validated LES for environmental applications against the Architectural Institute of Japan benchmark cases, while Antoniou et al. (2019) and Kim and Lee (2024) assessed CFD performance against field measurements from weather stations in dense urban configurations. The above studies illustrate the level of accuracy that well-executed simulations can achieve. Although the accuracy of CFD for environmental applications has been well established, a single urban CFD case of the scale needed for practical microclimate assessment, that typically spans several square kilometers with building-resolving resolution, requires tens to hundreds of CPU or GPU hours. This makes CFD unsuitable for the rapid scenario assessments demanded by urban design optimization, early-stage planning and multi-directional wind analyses. CityFFD, the fast fluid dynamics solver used in this work, partially addresses this computational burden through its GPU-based, semi-Lagrangian architecture (Mortezazadeh et al., 2022), although even with this approach approximately eight hours per case are needed at the spatial scales considered here. Therefore, there is a clear engineering need for surrogate models that can produce accurate microclimate predictions at a fraction of this computational cost.

The rapid development of deep learning during the past decade has opened new possibilities for accelerating fluid dynamics simulations (Goodfellow, 2016; Karniadakis et al., 2021). Within wind engineering and urban microclimate, several data-driven approaches have been proposed to replace or accelerate CFD. Zhao et al. (2024) proposed a machine learning surrogate to assess the impact of a new building on the local urban microclimate at the design stage, demonstrating that well-trained models can produce wind velocity and temperature estimates in seconds. Wang et al. (2025) evaluated a deep learning surrogate for predicting wind distribution in urban microclimate design, reporting good agreement with RANS for a range of morphological configurations. Han et al. (2024) used deep learning combined with land use data for the spatio-temporal prediction of outdoor microclimate, while Fujiwara et al. (2024) combined street-level and satellite imagery within a multimodal framework for climatic parameter prediction. Also, microclimate prediction from street-view-based descriptors and image data has received growing attention, as illustrated by Klimenka et al. (2025). More broadly, physics-informed machine learning approaches have been explored in order to incorporate conservation-law constraints directly into neural network training (Raissi et al., 2019), and recent advances in neural operators have offered a mathematical framework for learning mappings between function spaces rather than between fixed-dimensional vectors (Azizzadenesheli et al., 2024; Lu et al., 2021). Among these, the Fourier neural operator (FNO) has attracted particular interest due to its resolution-invariant properties and efficiency in solving parametric partial differential equations (Li et al., 2021), with applications demonstrated in weather prediction (Pathak et al., 2022; Bi et al., 2023; Lam et al., 2023), turbulent flow modeling, plasma surrogate modeling (Gopakumar et al., 2024), and seismic wave simulation (Li, B. et al., 2023).

Regarding urban microclimate specifically, Peng et al. (2024) applied FNO to three-dimensional transient urban wind simulation using CityFFD-generated data, achieving real-time prediction at 25 times the speed of the GPU-based solver. A more significant step was taken by Qin et al. (2025a), who introduced the localized Fourier neural operator (Local-FNO) for time-series prediction of multivariable, high-resolution 3D urban microclimate fields. Local-FNO addresses three key limitations of standard FNO, namely blurry output quality, high GPU memory demand, and large data requirements, by combining local patch-based training with geometry encoding and patch-overlapping techniques. Trained on data from a limited number of urban configurations, Local-FNO achieved mean absolute errors of 0.35 m/s for velocity and 0.30 °C for temperature and operates at nearly 50 times the speed of CFD. An earlier version of the present study was released as a preprint (Qin et al., 2025b). Even more recently, Lu et al. (2025) proposed a machine learning approach based on the UrbanTALES large-eddy simulation dataset to predict pedestrian-level wind fields from urban morphology and prevailing wind direction, reporting that the model recovers approximately 90% of CFD-level performance across idealized and realistic urban configurations. The above studies collectively establish that geometry-aware, locally trained neural operators represent a promising direction for rapid urban microclimate prediction.

Although the above progress is encouraging, several challenges remain that create research gaps in the current state-of-the-art. The generalization of data-driven models to urban configurations that are entirely absent from the training dataset, both in terms of morphology and spatial scale, has not been systematically demonstrated. Most published frameworks are validated on configurations that share characteristics with their training sets, and the question of how prediction accuracy degrades when the model encounters genuinely novel morphologies has not been thoroughly addressed. Also, directly predicting three-dimensional time-averaged wind and temperature fields from building geometry over urban areas exceeding 1 $km^2$ remains challenging with limited training data. It is worth noting that from an engineering standpoint, virtually none of the existing fast microclimate prediction frameworks provide a quantification of the confidence associated with their predictions. This is an important gap, since in practice engineers and urban planners not only need fast estimates of wind and temperature fields but also need to understand where those estimates can be trusted and where additional scrutiny is warranted. The above is particularly relevant for applications such as pedestrian wind assessment, urban heat island mitigation, and building energy analysis, where the consequences of inaccurate predictions carry real design and safety implications.

To close the above research gaps, this paper presents the Fast Localized Urban Microclimate Emulator (FLUME). The core assumption of the framework is that, under prescribed inflow and thermal forcing, much of the spatial variation of the mean wind and temperature fields is controlled by the surrounding building geometry. This is encoded through the Multi-Directional Distance Feature (MDDF), which captures directional distances to buildings via visibility-based manner, over adaptive spatial scales across 360°. By decomposing large CFD domains into local 3D patches, FLUME substantially increases the effective number of training samples from only 23 urban cases, to enable robust learning under data-limited conditions. The model simultaneously predicts the three components of mean wind velocity and mean temperature in full three-dimensional fields. Validation is conducted against five unseen urban configurations, while for two of them more detailed results are presented, namely a 1.2 km × 1.2 km case located in Montreal, Canada and a larger 3 km × 3 km case in Osaka, Japan, using physical validation metrics established in the wind engineering and microclimate literature (Franke et al., 2011; Tominaga et al., 2008; Okaze et al., 2021). FLUME produces a full 3D prediction in 10 seconds after training, with total prediction time remaining under one minute when MDDF computation is included. Besides the validation, an uncertainty quantification (UQ) framework is introduced based on deep ensembles (Lakshminarayanan et al., 2017) combined with conformal prediction (Shafer and Vovk, 2007; Angelopoulos and Bates, 2021), that provides prediction intervals over the predicted fields and enables local assessment of prediction reliability. The present study focuses on generalization across urban morphology and spatial extent under fixed inflow and thermal forcing. Generalization across inflow speed and direction, atmospheric stability, thermal boundary conditions, and mesoscale states is outside the present scope.

The paper is organized as follows. Section 2 describes the CFD dataset generated by CityFFD, the FLUME framework and its key components including the MDDF, the model training procedure, and the validation and uncertainty quantification methodology. Section 3 presents and discusses the results for the two unseen urban configurations and the UQ framework. Section 4 identifies the current limitations of the framework and outlines directions for future work. Section 5 summarizes the main conclusions. All the 3D datasets and a live interactive demo of the FLUME model are made available here:

https://flume.cityffd.ai/ and

https://huggingface.co/datasets/UMBE-LAB/CityFFD-UrbanMicroclimate

https://zenodo.org/records/22931677

# 2. Methods

## 2.1. Numerical dataset

Modelling wind flow and temperature in the lower ABL has seen major scientific and industrial interest during the last decades, while RANS and LES methods have been extensively studied for various environmental and structural applications (Potsis et al., 2023). In this study, an in-house LES code based on a semi-Lagrangian approach, the so-called CityFFD (Mortezazadeh et al., 2022), is utilized. Its core implementation is done in CUDA-C++ for GPU execution and is designed to resolve wind and temperature dynamics in urban environments. CityFFD can also be coupled with mesoscale models (Mortezazadeh et al., 2021); for instance, Wang et al. (2023) demonstrated a 50% reduction in error when replacing an empirical inflow with a WRF-informed one. Validation studies of CityFFD are presented in Mortezazadeh and Wang (2020) and Mortezazadeh et al. (2022) and more recently in Peng et al., (2024) based on data from the AIJ dataset. In the present paper, CityFFD serves as the ground truth for training and validating FLUME, thus further details on the solver are not included.

Figure 1a presents the computational domain used to generate the numerical dataset of 31 urban wind and temperature fields, each spanning 1.2 km × 1.2 km. The computational domain dimensions were estimated following RANS best practice guidelines (Tominaga et al., 2008), and all analyses were conducted in full-scale dimensions. The mesh is refined near the buildings with a structured resolution of 4 m × 4 m × 1.5 m in the along-wind, cross-wind, and vertical directions, respectively – see Figure 1b. At the inflow, an open terrain power-law profile is applied with α = 0.15 and U = 4 m/s at 10 m height. No-slip conditions are assigned to ground and building surfaces, symmetry conditions to the lateral and top boundaries, and a constant pressure condition at the outlet. Building surfaces were maintained at 40 °C, the inlet air temperature was set to 30 °C, and radiation effects were not modelled. A time step of 0.5 s was selected, yielding a Courant–Friedrichs–Lewy (CFL) number of approximately 0.5. Each simulation was run for a full-scale duration of 2.7 hours, with time-averaged wind and temperature quantities extracted after the first hour of spin-up. Each case utilized a single GPU for approximately 8 hours on a computational domain of 14 million cells. In addition, two urban configurations of 3 km × 3 km were included in the dataset, using the same grid resolution as the 1.2 km cases to ensure consistency across the dataset.

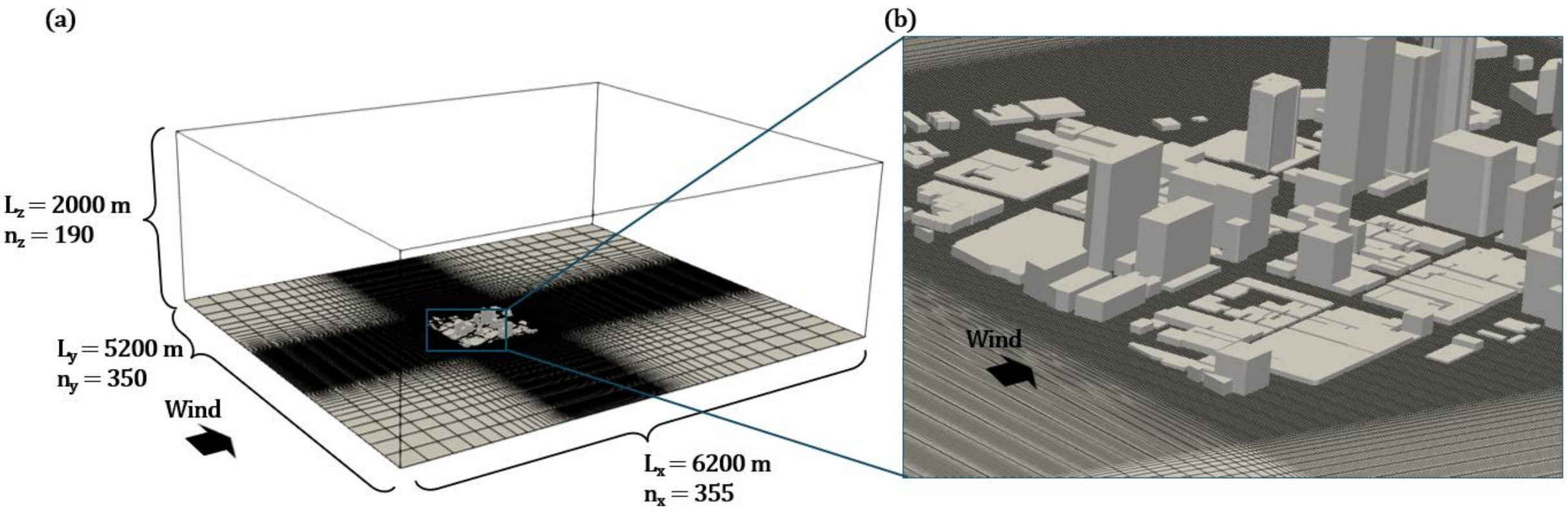

Figure 1: (a) Computational domain of CityFFD cases, (b) mesh near the buildings

The dataset comprises urban blocks from real cities worldwide, including London, Montreal, Tokyo, Singapore, and Hong Kong, among others. The 33 urban configurations were selected based on urban morphology indicators (Biljecki and Chow, 2022), with particular focus on building footprint area, height, volume, and shape complexity metrics. These descriptors have been shown to correlate closely with urban microclimate and wind flow patterns (Hu et al., 2024; Kaseb et al., 2020). Preference was given to blocks containing relatively tall buildings to include more complex three-dimensional wind fields and provide a more challenging and meaningful assessment of predictive performance. Although the above indicators provide a reasonable basis for the selection, the relationship between the morphological diversity of a training dataset and the generalization ability of a surrogate model trained on it is still an open research question. No established selection criterion exists at this stage, and the scope of this paper is not to identify the optimal set of urban morphologies for training purposes. Rather, the target is to quantify the generalization ability of FLUME across a range of realistic urban configurations, thus the selection of the 33 cases was not studied further in this respect. In Figure 2 the building height maps for all cases in the CityFFD dataset (Cases 0 to 32) are presented, with wind applied from left to right in all cases. Details on how these data were partitioned for training, testing, and validation of FLUME can be found in Section 2.2.

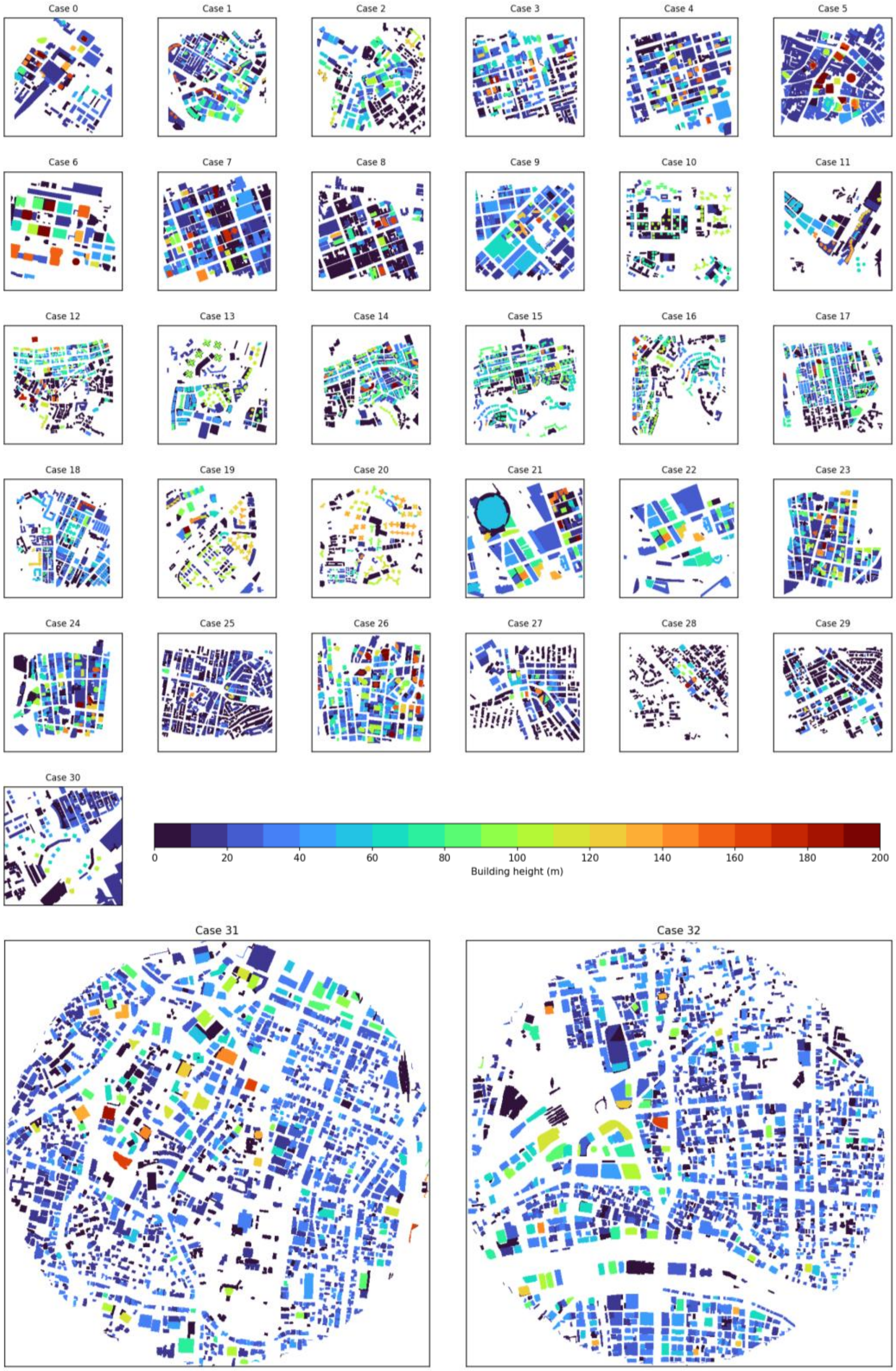

Figure 2: Urban configurations in our CityFFD dataset colored with building height

## 2.2. FLUME

### 2.2.1 Motivation

The goal of this study is to train a data-efficient model that can directly predict time-averaged three-dimensional urban airflow and temperature fields from 3D building geometry. This is challenging because a high-resolution urban CFD field contains millions of grid points. Directly learning a mapping from the full urban geometry to the full 3D flow field would therefore require a large amount of CFD data, while such data are expensive and slow to generate. To reduce the data requirement, the model must use the physical structure of the problem. In an urban area, the flow at a given location is strongly related to the shape of the open space around that location. This open space may be a wide plaza, a street canyon, or a narrow gap between buildings. What matters is not only the nearest building distance, but also how far the air space extends in different directions before being blocked by surrounding buildings. We describe this surrounding open region as the line-of-sight free space. It is the unobstructed space that can be seen from a point before straight rays hit building surfaces. Although airflow itself is not restricted to straight rays, this visibility-based description provides a simple and physically meaningful way to describe the geometric space that controls local blockage, channeling, acceleration, and wake formation.

The key difficulty is that line-of-sight free space has no fixed size or shape. A fixed local patch may cover only part of a large open square, or it may fully contain a narrow passage between buildings. Therefore, instead of trying to manually divide the city into physically meaningful open-space regions, we first convert the building geometry into a pointwise geometric feature field. This feature field describes the visible open space structure around each grid point and can then be cropped into standard 3D patches for model training. Based on this idea, we propose FLUME that consists of two main components: the Multi-directional Distance Feature, which encodes the line-of-sight free space around each point, and a local FNO model enhanced with gated MLP, which predicts the local velocity and temperature fields from the encoded geometric features. The complete workflow is shown on the left side of Figure 3. Starting from a building occupancy grid, FLUME first performs global geometry encoding, then divides the encoded fields into local 3D patches, applies the same local prediction model to all patches, and finally stitches the patch predictions into the full steady-state microclimate field. The dimensions marked in Figure 3 use a 300 × 300 × 160 domain as an example.

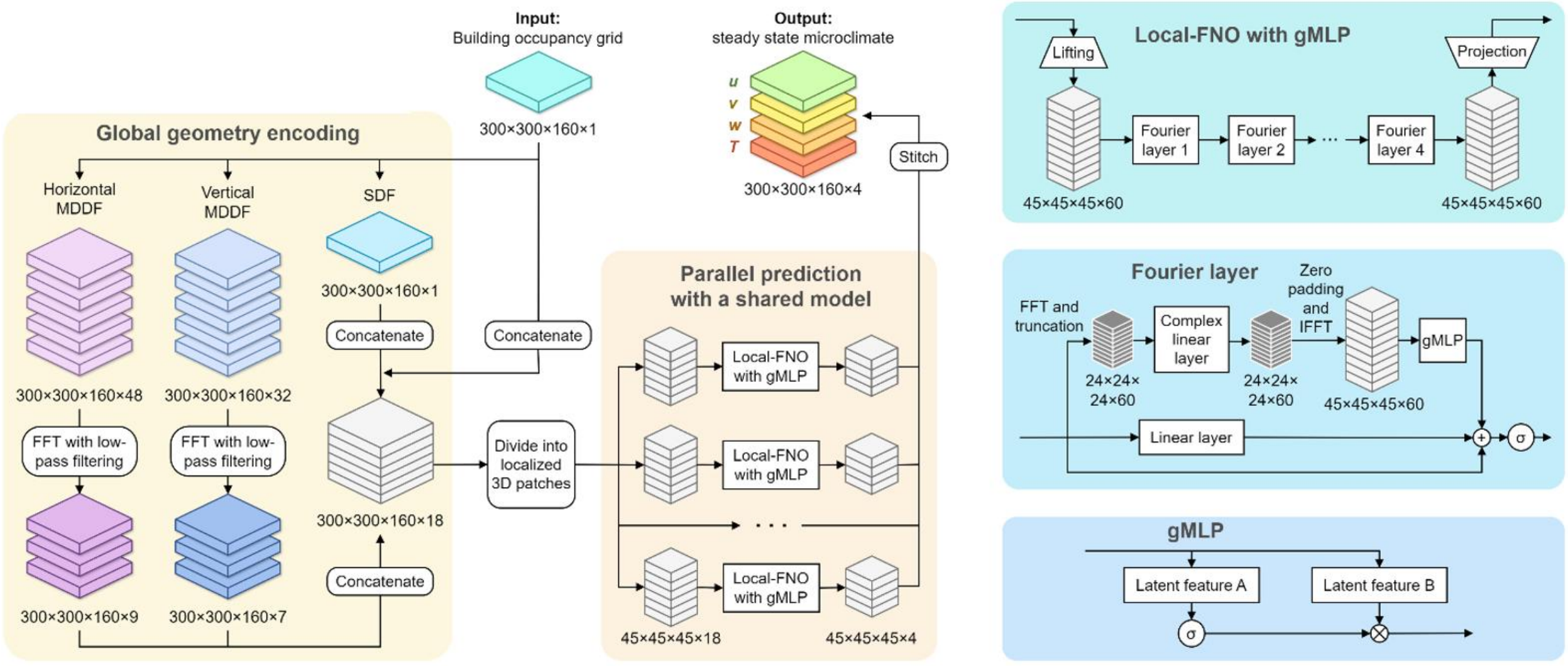

Figure 3: FLUME framework and gMLP-enhanced Local-FNO architecture

The overall strategy is therefore: global geometric encoding first, local flow prediction second. The geometry is encoded using the full urban domain, and the neural network is trained locally on patches extracted from the encoded fields.

### 2.2.2 Multi-directional Distance Feature

To encode the line-of-sight free space, we introduce the Multi-directional Distance Feature. For each grid point in the air domain, MDDF measures the distance to the first building obstacle along a set of predefined directions. A long distance in one direction means that the space is open in that direction, while a short distance means that the point is quickly blocked by a nearby building. Together, these directional distances describe the size, orientation, and shape of the visible open space around the point. MDDF is computed from the full-domain building occupancy grid before any patch extraction, and this order is essential to the framework. If the geometry were cropped first, the patch boundary would artificially cut off the visible space. In our method, a building outside a later training patch can still be encoded in the MDDF of a point inside the patch, as long as that building is directly visible along one of the sampled directions. In other words, the model does not only see the raw buildings inside a patch; it sees a precomputed description of the visible free space around every point in that patch.

In this study, MDDF is calculated in two complementary planes. First, it is computed in the horizontal plane to capture plan-view spatial structures such as plazas, street canyons, lateral blockage, and gaps between buildings. Second, it is computed in the vertical plane parallel to the incoming wind direction. This wind-aligned vertical MDDF captures the vertical obstruction and opening structure along the main flow path, including the effect of building height on upstream blockage, downstream wake development, and vertical flow redistribution. The first row of Figure 4 illustrates this idea using several directional MDDF fields. Compared with SDF, which only gives the nearest distance to buildings, each MDDF field shows how far the open space extends along one specified direction.

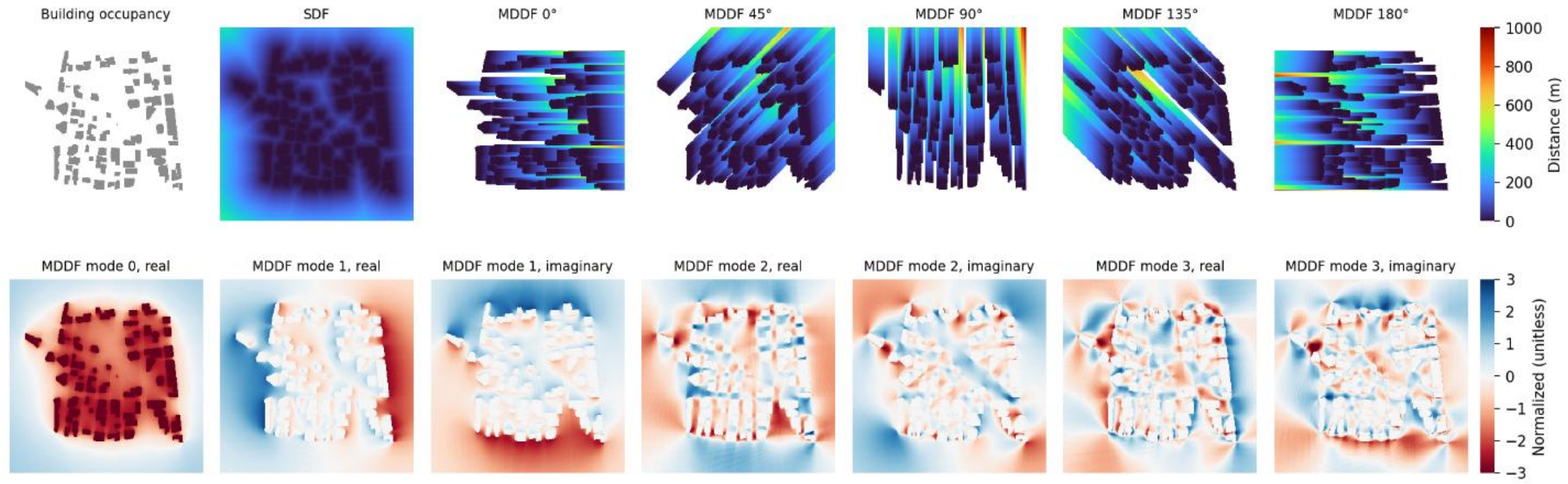

Figure 4: SDF and MDDF geometric feature visualization

The directional distances are sampled at multiple angles. However, directly using all angular samples would increase the number of input channels and may introduce high-frequency noise caused by grid discretization. Therefore, after MDDF is computed, a Fourier transform is applied along the angular dimension. Only the low-frequency angular components are retained. These low-frequency components preserve the main shape of the visible free space while reducing the input size and filtering out small angular fluctuations that are less relevant to the time-averaged flow. The second row of Figure 4 shows examples of the retained low-frequency MDDF components. These real and imaginary components still respond clearly to the direction and position of surrounding buildings, showing that the compressed MDDF fields preserve the main directional structure of the urban open space.

The final geometric input to the model consists of the compressed MDDF, the signed distance function, and global coordinates. The signed distance function provides the nearest obstacle distance, while MDDF provides directional visibility-based information about the surrounding open space. These features complement each other: SDF tells the model how close the nearest building is, and MDDF tells the model how the free space is distributed around the point.

**2.2.3 Local training with gMLP-enhanced Local-FNO**

After the full-domain geometric feature fields are computed, they are divided into local 3D patches. Each patch contains the compressed MDDF, SDF, and coordinate features for all grid points inside the patch. The corresponding CFD velocity and temperature fields inside the same patch are used as the training target. This local training strategy turns each full CFD simulation into many local training samples. Although the number of complete CFD cases is limited, each 3D field contains many different local spatial configurations, such as open regions, blocked regions, street canyons, and wake zones. Since the same fluid and heat transfer principles apply throughout the domain, the model can learn a common relationship between local geometric space and local flow response from these many patches.

The patch-based training does not mean that the model ignores buildings outside the patch. The important geometric context has already been encoded before patch extraction through MDDF. Therefore, the patch is only the computational unit used by the neural network; it is not the boundary of the geometric information used to describe the flow environment.

The local prediction model in FLUME is a Local-FNO, which applies FNO to local 3D patches, enhanced with gated MLP. Fourier neural operators are designed for field-to-field prediction and can efficiently combine spatial information over a grid by working with smooth spatial patterns in the Fourier domain. In this work, the FNO is applied to local 3D patches rather than to the full urban domain, allowing the model to learn repeated local flow patterns from limited CFD data. The right side of Figure 3 shows the local prediction model used in each patch. The input patch is first lifted to a higher-dimensional feature space, processed by four Fourier layers with gMLP, and then projected to the output variables.

To better represent the relationships between different input feature channels, the standard MLP component in the FNO block is replaced by a gated MLP, or gMLP. The gating mechanism allows the model to adaptively combine information from different channels, such as coordinates, SDF, and MDDF components. This is intended to improve the model's ability to learn how different geometric descriptors jointly affect the predicted velocity and temperature fields. The lower part of Figure 3 illustrates the Fourier layer and gMLP structure. The Fourier layer combines spatial information within the patch, while gMLP further controls how latent feature channels interact.

During inference, the encoded full-domain feature fields are divided into overlapping patches. Local-FNO predicts the flow and temperature fields for each patch, and the patch predictions are assembled to recover the full 3D field. Overlapping patches are used so that the final prediction can rely more on the central region of each patch, where boundary effects are smaller.

For applications with different incoming wind directions or reference wind speeds, the geometry and predictions can be adjusted accordingly. A different wind direction can be accommodated by rotating the urban geometry to align the incoming wind with the model's reference direction before feature encoding, then transforming the predicted fields and horizontal velocity components back to the original coordinates. For a different reference wind speed, the predicted velocity components can be approximately rescaled by the ratio of the target to the training reference wind speed, assuming Reynolds-number independence, an unchanged inlet profile shape, and negligible buoyancy effects.

#### 2.2.4 Practical advantages of FLUME

Beyond data efficiency, the FLUME design also provides several practical advantages for large-scale urban microclimate prediction. Because the model is trained and applied on local patches, it greatly reduces the GPU memory required to process high-resolution 3D CFD data. This makes it possible to train on large urban airflow datasets using relatively small GPUs. The same advantage also applies during inference: local patches can be predicted independently and in parallel, while each prediction only requires a limited amount of memory. The local formulation also removes the requirement that all CFD cases have the same full-domain size. As long as the extracted 3D patches have the same physical size and grid resolution, FLUME can be trained using CFD data from urban domains of different overall scales. After training, the model can also be applied to urban areas of arbitrary size by computing the geometric feature fields and dividing them into standard local patches.

Finally, local prediction provides flexibility for practical applications. It is not always necessary to predict the entire 3D wind and temperature field. For design analysis or building optimization, the user may only need the microclimate around a specific building, street canyon, public space, or area affected by a proposed design change. In such cases, FLUME can predict only the selected local regions, further reducing computational cost while focusing the prediction on the area of interest.

**2.2.5 Training details**

In many practical applications, only a limited number of high-resolution three-dimensional urban CFD cases are available, and generating such simulations is computationally expensive. The ability to produce useful predictions from a small number of training cases is therefore important for the practical applicability of surrogate models. In this study, each data split uses 23 urban CFD cases for training, 3 for validation, and 7 for testing, as shown in Figure 12. The results demonstrate that FLUME can produce meaningful predictions for unseen urban configurations using only these 23 training cases. This level of data efficiency is notable because the training cases can be real complex urban areas with dense buildings, irregular layouts, and large height variations, rather than simplified benchmark cases consisting of only a few box-shaped buildings. In the present implementation, the training data include urban domains with grid sizes of 300 × 300 × 160 and 750 × 750 × 120. All samples are extracted as local 3D patches of 45 × 45 × 45 grid cells. This unified patch size allows CFD cases with different full-domain sizes to be used together during training. The horizontal MDDF uses 48 angular directions, and the wind-aligned vertical MDDF uses 32 angular directions. After applying FFT along the angular dimension, the retained low-frequency components produce 9 horizontal MDDF channels and 7 vertical MDDF channels. These channels are used as part of the model input together with SDF and global coordinates. The same channel reduction is also illustrated in Figure 3, where the 48 horizontal MDDF directions and 32 vertical MDDF directions are compressed into 9 and 7 channels, respectively, before being concatenated with the other input features.

The local FNO uses 4 layers, a hidden width of 60, and 12 Fourier modes in each spatial direction. The standard MLP in the FNO block is replaced by gMLP. The training loss is the root mean square error (RMSE) between the predicted fields and the CFD reference fields. Training takes approximately 6 hours on a single NVIDIA V100 GPU with 32 GB memory. For a 300 × 300 × 160 urban domain, full-domain prediction takes about 10 seconds. If a larger patch overlap is used to improve the smoothness of the reconstructed field, the inference time increases accordingly. In general, the inference time can be reasonably approximated to be proportional to the total number of grid cells being predicted.

### 2.3 Validation and Uncertainty Quantification

The validation of FLUME is conducted by comparing its predictions against the CityFFD reference solution for two unseen configurations. In Sections 3.1 and 3.2, predictive accuracy is assessed using patch-averaged metrics computed over the full set of patches. Let $\hat{y}_i$ denote the predicted value and $y_i$ the corresponding reference value at point $i$, over a total of $N$ sample points. The metrics include the coefficient of determination $R^2$ (Equation 1), the Root Mean Square Error RMSE (Equation 2), and the mean absolute error MAE (Equation 3). Additionally, the validation approach follows the methodology proposed by Franke et al. (2011) and Tominaga et al. (2008) for the assessment of physics-based models such as RANS and LES against wind tunnel results. In these guidelines, the Factor of 2 (FAC2) is defined as the fraction of data points for which the predicted values fall within a factor of 0.5 to 2 of the reference values (Equation 4), where W was taken as 0.05.

$$R^2 = 1 - \frac{\sum_{i=1}^{N}(y_i - \hat{y}_i)^2}{\sum_{i=1}^{N}(y_i - \bar{y})^2} \qquad \text{Equation 1}$$

$$RMSE = \sqrt{\frac{1}{N}\sum_{i=1}^{N}(y_i - \hat{y}_i)^2}$$ Equation 2

$$MAE = \frac{1}{N}\sum_{i=1}^{N}|y_i - \hat{y}_i|$$ Equation 3

$$FAC2 = \frac{1}{N}\sum_{i=1}^{N} n_i\,, with\; n_i = \begin{cases} 1\; for \left[0.5 \leq \frac{\hat{y}_i}{y_i} \leq 2.0\right] or\; [\hat{y}_i \leq W\; and\; y_i \leq W] \\ 0\; else \end{cases}$$ Equation 4

In recent studies validating LES against the AIJ wind tunnel dataset for buildings in idealized urban configurations (Potsis et al., 2024), FAC2 at pedestrian height ranged from 0.75 to 0.93, well above the threshold of 0.5 proposed by Okaze et al. (2021) as the minimum acceptable performance for urban wind simulations. Similar comparisons between CFD and real weather stations can be found in Kim and Lee (2024), who validated a CFD model of a dense urban area against measurements during daytime heatwaves, reporting $R^2$ values of 0.96 to 0.98 for temperature and 0.64 to 0.81 for wind speed across five locations. The above results illustrate the level of agreement that physics-based models can achieve against field data. At the same time, it is clear that the scatter associated with wind speed predictions in realistic urban environments is inherently greater compared to controlled laboratory conditions, which is expected due to the fact that the complexity of the flow field in real urban areas is significantly higher.

Beyond these metrics, the validation presented uses accuracy thresholds for variables of engineering interest. Based on recent extensive research in urban microclimate and wind engineering, thresholds of 0.5 m/s for velocity components and 1 °C for temperature are adopted as the MAE values below which a prediction is considered acceptable for practical applications in pedestrian wind assessment, thermal comfort analysis, and urban energy modeling (Yang et al., 2023; Antoniou et al., 2019; Brozovsky et al., 2022; Hong et al., 2021; Sezer et al., 2023).

In addition to validating the accuracy between FLUME and the CityFFD reference solution, the confidence in the predictions is a critical aspect for practical engineering use. Uncertainty quantification (UQ) in neural networks has become a major research topic in recent years (Nguyen et al., 2021; Gawlikowski et al., 2023), since fast prediction frameworks have limited practical impact in engineering applications if their outputs cannot be accompanied by a meaningful estimate of reliability. Even when validation studies demonstrate good global accuracy, UQ can deepen the understanding of local prediction confidence and reveal where the model is most sensitive to the composition of the training data. In this paper, two complementary methods are employed to establish the UQ framework for FLUME. First, deep ensembles (Lakshminarayanan et al., 2017) are used as a practical proxy for epistemic uncertainty and as a measure of sensitivity to the specific training and testing data subsets. Let M be the number of models trained with various training/testing/validation subsets, μ the mean ensemble prediction of all M models (Equation 5) and σ the standard deviation of the predictions from M models (Equation 6), which serves as the practical proxy of the epistemic uncertainty (ensemble). Second, conformal prediction (Shafer and Vovk, 2007; Angelopoulos and Bates, 2021) is applied to transform the ensemble uncertainty estimates into calibrated prediction intervals for each spatial location, providing statistical bounds that complement the model's prediction. To achieve this  for a given case, a number of calibration points are selected from other unseen cases, and their normalized nonconformity score s is defined in Equation 7. Given a desired miscoverage level α, the conformal quantile q is defined in Equation 8, where Q is the quantile function and n is the number of calibration points. The resulting prediction interval $C_a$ is then defined in Equation 9. Further details on the number of ensemble members, unseen cases, and calibration points are discussed in Section 3.3. This straightforward two-stage process provides prediction intervals whose width adapts automatically to the local aerodynamic complexity of the flow, generating a spatially resolved and physical measure of confidence in the predictions of FLUME.

$$\mu(x) = \frac{1}{M}\sum_{m=1}^{M} \hat{y}^m(x) \qquad \text{Equation 5}$$

$$\sigma(x) = \sqrt{\frac{1}{M}\sum_{m=1}^{M} (\hat{y}^m(x) - \mu(x))^2} \qquad \text{Equation 6}$$

$$s_i = \frac{|y_i - \mu(x_i)|}{\sigma(x_i)}$$ Equation 7

$$\mathrm{q}(a) = Q\left(s_i; \frac{(1-\alpha)\cdot(n+1)}{n}\right)$$ Equation 8

$$C_a(x) = \ [\mu(x) - q(\alpha)\cdot\sigma(x), \mu(x) + q(\alpha)\cdot\sigma(x)\ ]$$ Equation 9

$$\hat{q}(\alpha) = s_{(\lceil(1-\alpha)(n_{cal}+1)\rceil)}$$

$$C_\alpha(x) = [\mu(x) - \hat{q}\alpha, \sigma(x), ; ; \mu(x) + \hat{q}\alpha, \sigma(x)]$$

## 3. Results and discussion

In this section, FLUME predictions for unseen configurations are compared with the CityFFD reference solutions, for a 1.2 km domain in Section 3.1 and a 3 km domain in Section 3.2. Furthermore, the proposed UQ framework is displayed and discussed in Section 3.3.

### 3.1. Results for a 1.2 km unseen urban configuration

The unseen configuration that was selected for the primary FLUME validation is Case 0, shown in Figure 2, which corresponds to a 1.2 km × 1.2 km area of downtown Montreal. This case was deliberately excluded from all training and validation data, meaning it contributes in no way to the learning process of the model. Therefore, the predictions of FLUME in this section are assessed against CityFFD results as a test of generalization to an unseen urban morphology.

In Table 1 the validation metrics for U (streamwise velocity), V (lateral velocity), W (vertical velocity), and T (temperature) are presented over all patches of this unseen configuration. As can be seen, the $R^2$ values indicate that FLUME reproduces the global spatial distributions of all four variables with strong agreement. For U and T, $R^2$ reaches 0.97, which confirms that the dominant aerodynamic and thermal patterns of the unseen case are captured with high fidelity by the model. Regarding the secondary velocity components, $R^2$ equals 0.86 for V and 0.83 for W. Although lower, these values still indicate that the main three-dimensional flow structures are recovered, despite the fact that these components are more sensitive to localized flow separation, recirculation, and complex building wake interactions. This hierarchy of predictive accuracy, highest for U and T, progressively lower for V and W, is physically consistent and has been observed in validation studies of LES in isolated buildings (Tominaga et al., 2008a; Okaze et al., 2021), idealized urban configurations (Potsis et al., 2024) and data-driven models (Qin et al., 2025a; Wang et al., 2025) for urban wind flows.

The MAE values are 0.20 m/s for U, 0.05 m/s for V, 0.09 m/s for W, and 0.19 °C for T, all of which remain well below the practical accuracy thresholds discussed in Section 2.3. The RMSE values follow the same trend, with the largest magnitudes found for U and T. This is consistent with the broader spatial variation and larger absolute values of these two quantities in the urban domain. Regarding the FAC2, the values are also high, reaching 0.98 for U, 0.90 for V, 0.83 for W, and 1.00 for T. To put the accuracy of these results into perspective, the FAC2 value for U is close to the upper end of the range reported by Potsis et al. (2024) for LES validation against AIJ wind tunnel data (0.75 to 0.93). The high FAC2 indicates a useful level of agreement with CityFFD despite the limited training data. Regarding T, the perfect FAC2 reflects the relatively smooth thermal field characteristic of the chosen boundary conditions, where building surface temperatures impose a coherent thermal forcing that FLUME captures consistently. The lower FAC2 for W is expected, due to the fact that vertical velocity components are weaker in magnitude and more sensitive to local three-dimensional building interactions, although it is still well above the 0.5 threshold of Okaze et al. (2021).

***Table 1.*** *Validation metrics for the unseen 1.2 km case (Case 0)*

| **Parameter** | **R2** | **RMSE** | **MAE** | **FAC2** |
|---|---|---|---|---|
| **U** | 0.97 | 0.33 | 0.2 | 0.98 |
| **V** | 0.86 | 0.07 | 0.05 | 0.90 |
| **W** | 0.83 | 0.13 | 0.09 | 0.83 |
| **T** | 0.97 | 0.33 | 0.19 | 1 |

Visualizing the flow field in terms of streamlines is equally important for understanding the prediction of FLUME and comparing it with the CityFFD reference solution. In Figure 5a the streamlines on a horizontal cross-section of the three-dimensional predicted and CityFFD flow fields at z = 10 m in full scale are presented, which is a reference height commonly adopted for inner boundary layer development (ASCE 49-21). The streamlines were generated by uniformly seeding the horizontal plane and integrating the local mean velocity field. In general, the flow pattern predicted by FLUME seems to agree in qualitative terms with the CityFFD results. The wind is channeled between the building blocks in a similar manner, the recirculation structures formed in the wake of the larger obstacles are reproduced, and the main left-to-right flow is consistent between the two. FLUME captures well the deflection of the flow around the central high-rise cluster and the elongated building in the southwest part of the domain, where the local geometry strongly redirects the wind. The largest differences can be found in confined regions near sharp building corners, narrow passages, and immediately downstream of obstacles. In these regions, even small errors in the local velocity direction can modify the exact trajectory of individual streamlines. This is expected, since the flow in these locations is governed by abrupt separation and recirculation that are inherently more difficult to reconstruct. Similar discrepancies have been reported in LES validation studies for urban configurations, both for isolated buildings (Tominaga et al., 2008a; Okaze et al., 2021) and building clusters (Potsis et al., 2024).

In Figure 5b the streamlines on a vertical plane at the center of the domain are presented, where four distinct buildings are resolved. This view is useful to understand the ability of FLUME to reproduce the three-dimensional redistribution of the flow over and behind urban obstacles. As can be seen, the model captures the main acceleration above rooftop level, the upward deflection in front of the tall buildings, and the recirculation zones on their leeward side. Compared with the horizontal plane, differences are more visible in localized wake and shear-layer regions. This is consistent with the lower FAC2 obtained for W, since vertical velocity is more sensitive to these complex interactions as already discussed. Despite the above, the main vertical flow structures are reproduced with good accuracy, which shows that FLUME preserves not only low global error but also the main flow regimes of the unseen configuration.

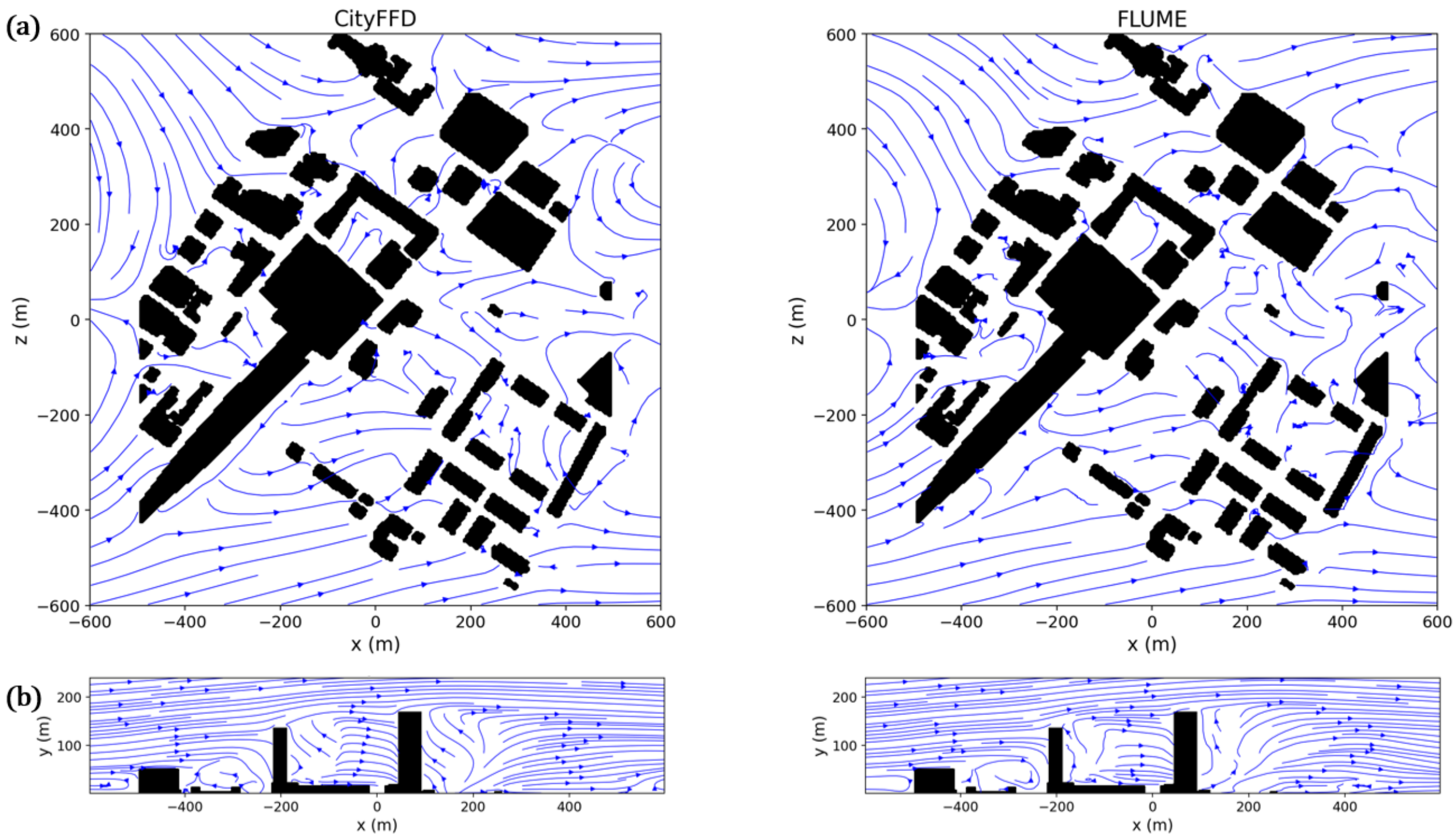

*Figure 5: Streamlines (a) on a horizontal plane at z = 10 m; (b) on a vertical plane at domain center*

In Figure 6 the full-plane comparisons of U, V, W, and T at z = 10 m are presented, with the CityFFD reference solution in the first column, the FLUME prediction in the second, and the difference field in the third. Regarding U, FLUME captures the overall distribution of the velocity, including the main high-velocity corridors in the open regions and the lower-velocity wakes behind building clusters. The sheltering effect of the central cluster and the elongated southwest building is consistently reproduced, which shows that the dominant flow organization is preserved. Local discrepancies are concentrated near sharp building edges, narrow passages, and immediately downstream of obstacles, where strong gradients make the exact reconstruction more difficult. It is worth noting that the majority of these differences remain below the 0.5 m/s threshold, as indicated by the white color in the difference colormap. Regarding V, the main positive and negative regions associated with flow deflection around corners and channeling inside the urban canopy are reproduced with good accuracy, particularly around the central building group. Localized deviations can be found near corners where the turning of the flow changes rapidly over short distances, which is expected due to the abrupt nature of the separation at these locations. Regarding W, FLUME reproduces the broad regions of upward and downward motion related to flow impingement, separation, and recirculation around the major obstacles. However, this component shows the largest local discrepancies among the velocity variables, which is consistent with what was discussed previously, since W is more sensitive to abrupt three-dimensional effects in the flow field. Finally, regarding T, the model captures the main thermal distribution including the warmer regions adjacent to building surfaces and the cooler ventilated zones. The residual differences are found mainly near building edges and in sheltered wake areas where sharp thermal gradients interact with local recirculation.

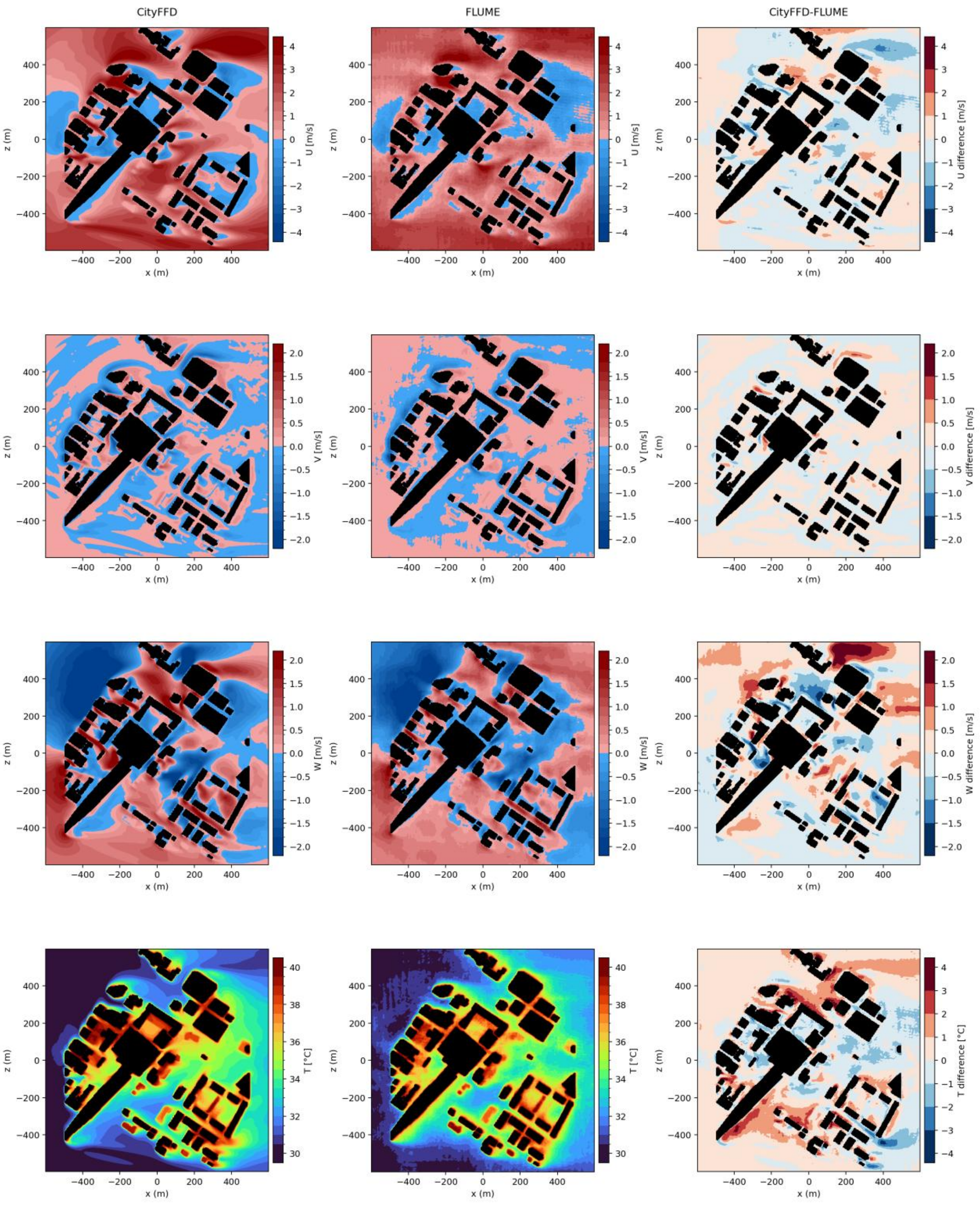

*Figure 6: Contour comparison of U, V, W, T at z = 10 m between CityFFD, FLUME, and their difference for Case 0*

In Figure 7 the corresponding comparisons on a vertical cross-section at the center of the domain are presented. Regarding U, the model captures the expected upstream deceleration, the acceleration above rooftop level, and the gradual downstream recovery. The most notable local differences are found near the roof-level shear layers and in the downstream recovery region, where separation and reattachment create sharp spatial gradients. This is consistent with what was observed in Figure 6 for the horizontal plane. Regarding V, FLUME reproduces the main positive and negative regions near the windward and leeward building faces, with residual discrepancies concentrated immediately downstream of the obstacles. Regarding W, the model recovers the main updraft and downdraft regions, although, similar to the horizontal plane, it exhibits somewhat larger differences near wake zones and around the tallest buildings. This is expected, since the vertical velocity component is governed by the most complex three-dimensional interactions in the flow field. Regarding T, the vertical thermal field is reproduced with very good accuracy, including the warmer air close to heated building surfaces and the smoother gradient further from the obstacles. The remaining differences are concentrated near buildings and in downstream regions where recirculation influences the temperature distribution.

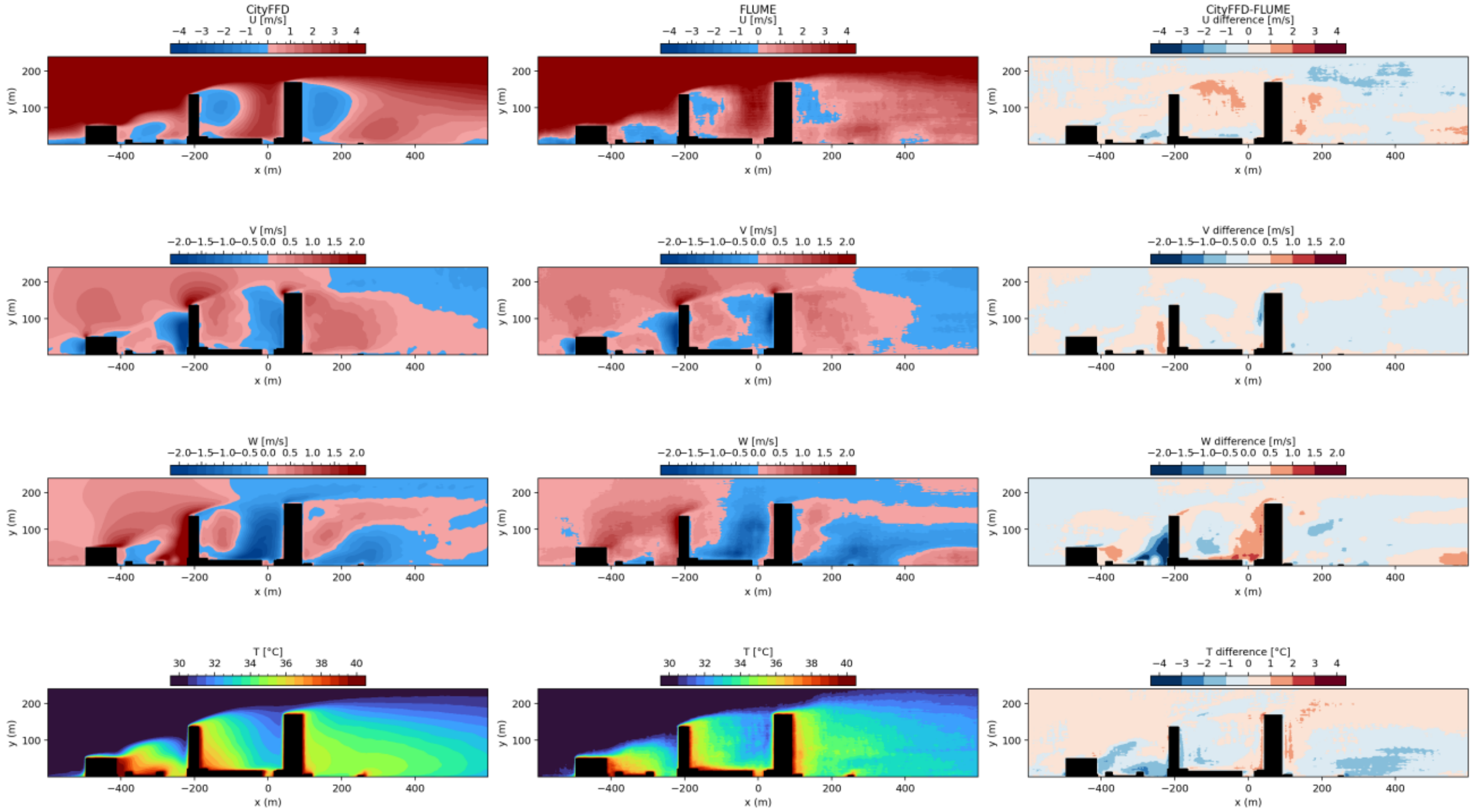

*Figure 7: Contour comparison of U, V, W, T at the middle of the domain between CityFFD, FLUME, and their difference for Case 0*

In Figure 8 the comparison between FLUME and CityFFD is summarized by plotting the values of U, V, W, and T for all patches of the unseen case. As can be seen, the points for U are tightly distributed around the 1:1 line, which confirms that FLUME reproduces very well the dominant streamwise component across patches with different urban characteristics. The scatter for V and W is naturally larger, which is consistent with what was discussed previously, since these two components are more sensitive to local effects such as cornering flow, wakes, separation, and recirculation. Despite that, the majority of the points remain close to the 1:1 line and within the bounds, which indicates satisfactory agreement for the secondary velocity components as well. Regarding temperature, the points follow very closely the 1:1 line and remain within the ± 1°C bounds. This confirms the strong predictive capability of FLUME for the thermal field. Overall, Figure 8 supports the quantitative metrics of Table 1 and the spatial comparisons of Figure 6 and Figure 7 and shows that FLUME maintains good agreement with CityFFD over a wide range of local aerodynamic and thermal conditions.

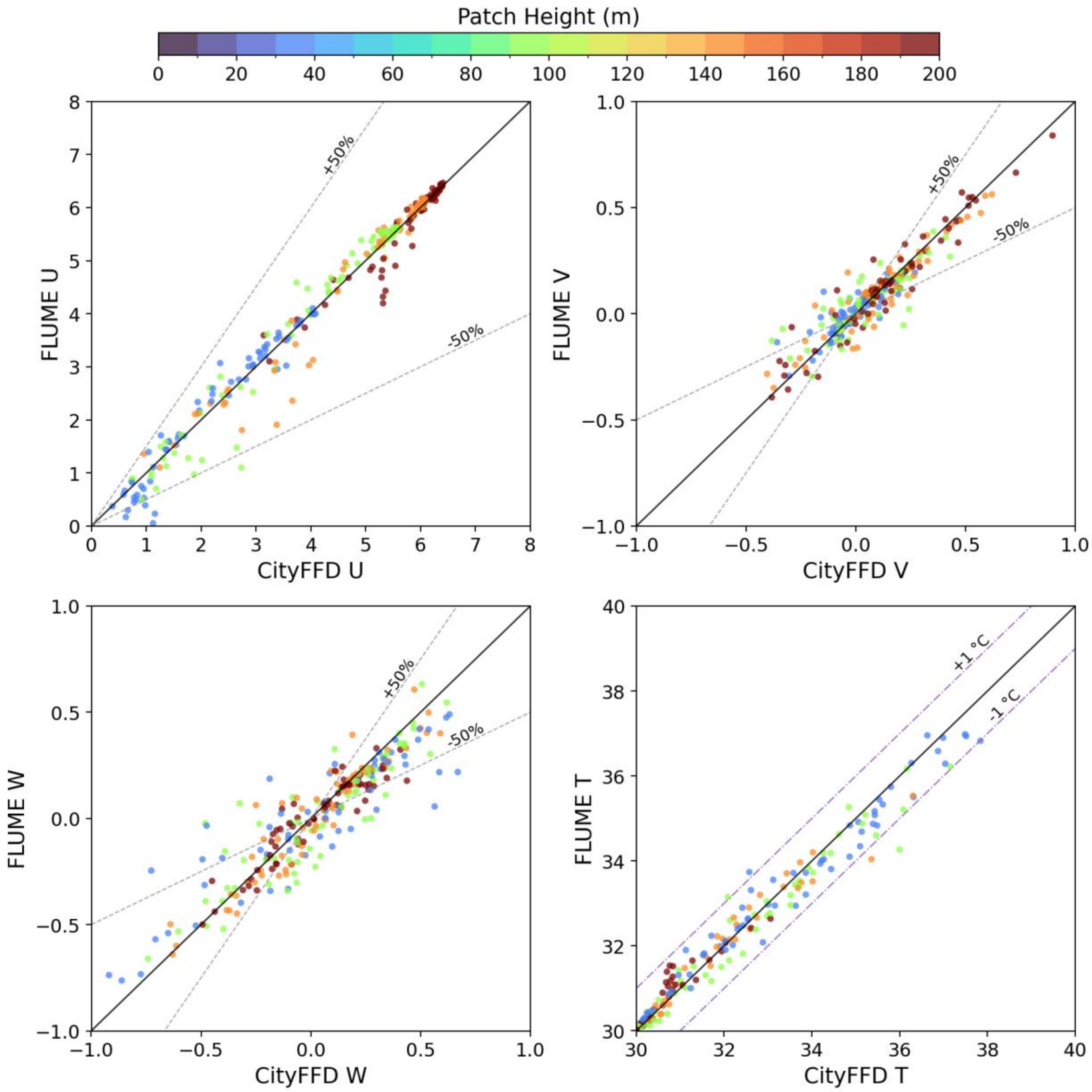

Figure 8: Correlation of U, V, W, T at patches from CityFFD and FLUME for Case 0

### 3.2. Results for a 3 km unseen case

The second unseen configuration used to assess FLUME corresponds to a larger urban area of 3 km × 3 km located in Osaka, Japan. This case is more demanding than the 1.2 km configuration, since it includes a substantially larger spatial extent, denser urban variability, and a wider range of aerodynamic interactions between building clusters. In Table 2 the validation metrics for U, V, W, and T are presented, based on the definitions given in Section 3.1. As can be seen, FLUME again reproduces the overall spatial distributions of all variables with strong agreement, particularly for U and T, with $R^2 = 0.95$ and $R^2 = 0.98$, respectively. This shows that the model preserves very well the dominant aerodynamic and thermal organization of the larger unseen configuration. Regarding the secondary velocity components, the agreement remains satisfactory, with $R^2 = 0.78$ for V and $R^2 = 0.70$ for W. This is consistent with the stronger sensitivity of these quantities to local turning of the flow, separation, wake interaction, and recirculation within a more complex urban canopy, as also discussed for the Montreal case. The MAE values are 0.31 m/s for U, 0.05 m/s for V, 0.09 m/s for W, and 0.18 °C for T, all of which remain below the practical thresholds discussed in Section 2.3. The FAC2 values are also high, reaching 0.99 for U, 0.86 for V, 0.79 for W, and 1.00 for T, which confirms that the majority of the predictions remain within acceptable agreement with the CityFFD reference solution. Compared with the smaller unseen case, the global performance is only slightly reduced for the velocity components. This is expected, since the larger domain contains a broader range of wakes, recirculation structures, and local aerodynamic interactions that are inherently more difficult to reconstruct. Overall, Table 2 indicates that FLUME generalizes well also to the 3 km unseen urban morphology and maintains strong predictive capability for all variables of interest.

***Table 2.*** *Validation metrics for the unseen 3 km case (Case 32)*

| Parameter | R2 | RMSE | MAE | FAC2 |
|---|---|---|---|---|
| **U** | 0.95 | 0.41 | 0.31 | 0.99 |
| **V** | 0.78 | 0.07 | 0.05 | 0.86 |
| **W** | 0.70 | 0.11 | 0.09 | 0.79 |
| **T** | 0.98 | 0.25 | 0.18 | 1 |

In Figure 9 the plan-view comparisons at z = 10 m for all four variables are presented. Regarding U, FLUME captures the large-scale organization of the flow, including the main high-velocity corridors in the open regions and the lower-velocity zones associated with sheltering and wake formation inside the dense urban core. The model also reproduces the broad curved bands of reduced U that extend through the urban interior. This is an interesting result, since it shows that coherent wake structures are preserved over longer spatial scales and not only within isolated local patches. The largest local discrepancies appear near the perimeter of the urbanized area and around concentrated building groups in the upper-left and lower parts of the domain, where wake structures are more diverse. Despite these deviations, the dominant flow pattern is not altered. Regarding V, FLUME reproduces the main deflection and channeling regions, especially in the central and upper parts of the domain. The most pronounced local differences can be found along oblique streets and around dense building clusters where the lateral redistribution of the flow depends strongly on subtle geometric details. Regarding W, the model captures the broad upward and downward motion bands around the major building clusters, although this component again shows the largest local differences, particularly in wall separation and recirculation regions. This is consistent with what was observed for the Montreal case. Regarding T, the agreement with CityFFD is very strong. Both the large warm core of the urbanized area and the cooler peripheral zones with smooth gradients toward the outer boundary are reproduced consistently, with residual discrepancies mainly near the outer edge of the domain and in sheltered regions of reduced ventilation.

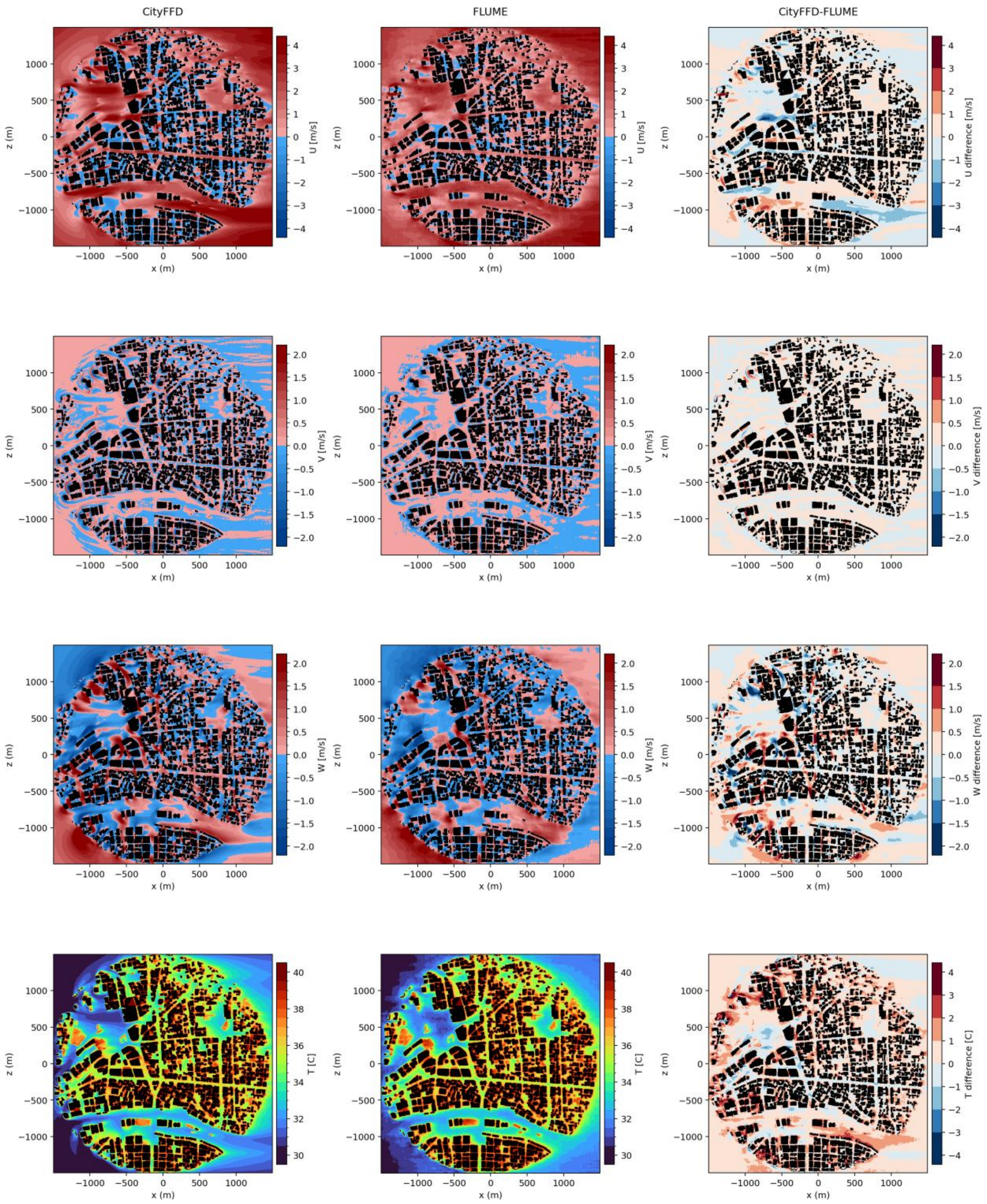

Figure 9: *Contour comparison of U, V, W, T at z = 10 m between CityFFD, FLUME, and their difference for Case 32*

In Figure 10 the comparison in a vertical plane located at the middle of the domain is presented. This view confirms that FLUME also preserves the main three-dimensional structures of the larger unseen case, although some limitations become more visible when the rooftop shear layers and wake recovery regions are examined in section. Regarding the streamwise velocity U, the model captures the expected upstream deceleration in front of the main obstacles, the acceleration above roof level, and the gradual recovery in the downstream region. The broad spatial organization is correct, but the difference map shows that FLUME tends to smooth some of the sharper transitions near rooftop level and immediately behind the taller obstacles, where separation and reattachment create narrow regions of strong gradient. This is physically reasonable, since such zones are governed by abrupt local changes that are more difficult to reconstruct exactly in a patch-based surrogate model. Regarding the lateral velocity component V, the agreement remains acceptable and the model reproduces the general positive and negative bands around the buildings. However, the field is less sharply resolved in places where the lateral motion changes rapidly close to the building faces. The local differences are again strongest near the main obstacles and in the immediate downstream region, which indicates that FLUME captures the sign and extent of the turning flow better than its exact local magnitude. Regarding the vertical velocity component W, the model reproduces reasonably well the broad updraft and downdraft regions, especially around the more prominent building elements. Still, as already observed in the horizontal plane, this is the most challenging variable, and the differences increase in regions where rooftop-induced vertical exchange and recirculation loops are strongest. Regarding temperature T, the vertical section shows very good agreement, particularly in the way FLUME captures the warmer air within the urban canopy and the cooler region aloft. The main remaining differences are concentrated close to the buildings and near the lower part of the section, where thermal gradients interact with local flow recirculation. This makes the temperature distribution more sensitive to small aerodynamic phenomena in these locations.

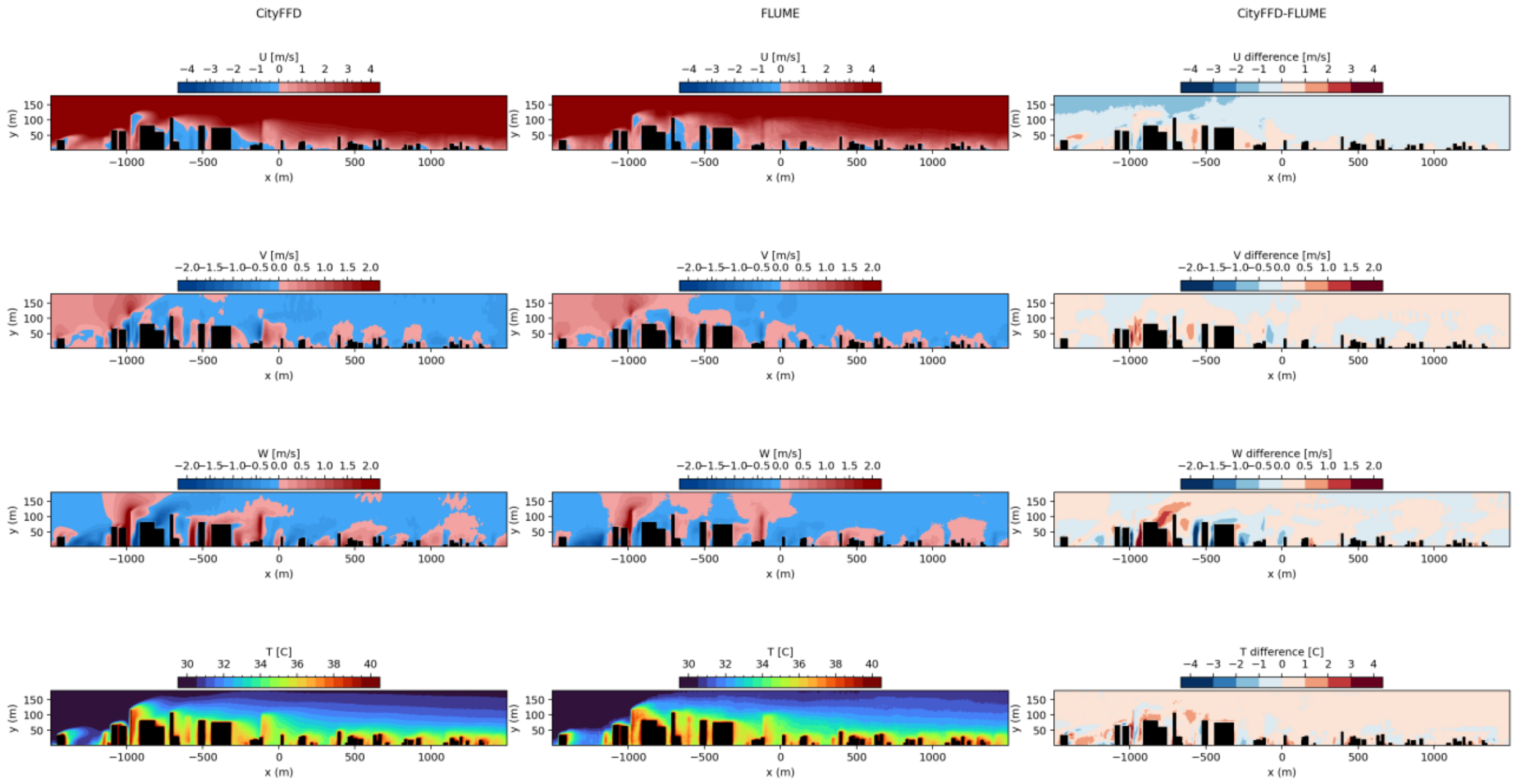

Figure 10: *Contour comparison of U, V, W, T at the middle of the domain between CityFFD, FLUME, and their difference for Case 32*

The correlation plots in Figure 11 provide a compact summary of the above observations across all patches of the unseen 3 km case. As can be seen, the scatter for U remains tightly aligned with the 1:1 line over the full range of values, which confirms that FLUME predicts the dominant streamwise flow robustly even as patch height and local morphology vary substantially. At the higher values, a modest spread becomes visible, with some points falling above or below the exact line. This is consistent with the local over- and under-predictions observed in the wake recovery and high-speed corridor regions of Figure 9. Regarding V, the correlation remains satisfactory, but the cloud is broader around the 1:1 line, reflecting the fact that lateral motions are weaker in magnitude and more sensitive to local turning around buildings. Regarding W, the scatter is broader still, with noticeably more dispersion at both positive and negative values. This agrees with the local differences observed in the wake and recirculation regions of both the horizontal and vertical planes. Despite that, the overall trend remains centered around the 1:1 line, which shows that FLUME still captures the dominant sign and magnitude range of the vertical motions. The temperature correlation is the strongest of all, with points following very closely the 1:1 line and remaining within the

± 1°C bounds across the range shown. Overall, similar to what was observed in the 1.2 km case of Section 3.1, Figure 11 supports the quantitative metrics of Table 2 and the spatial comparisons of Figures 9 and 10 and shows that FLUME maintains good agreement with CityFFD over a wide range of local aerodynamic and thermal conditions, even for the larger unseen urban configuration.

Furthermore, it is worth noting that this 3 km case is more demanding than the 1.2 km unseen case not only due to its larger spatial extent, but also due to the composition of the training dataset. Among the training cases, only one other city of similar area, namely Case 31 in Figure 2, provides patches extracted from a domain of comparable size, while the remaining training patches are obtained from the smaller 1.2 km cases. This means that the model is required to generalize to a broader-scale urban organization with limited direct exposure to similarly large spatial structures during training. To assess the contribution of this large-area training case, a separate model was trained without Case 31. Its prediction accuracy for the unseen Case 32 decreased by approximately 20% for all liner regression metrics, compared with the model trained with Case 31 included. This indicates that, although the patch-based formulation allows FLUME to learn from local flow and geometry relationships across domains of different size, including at least one training case with a comparable total area improves the ability of the model to reconstruct the larger-scale aerodynamic and thermal organization of unseen configurations.

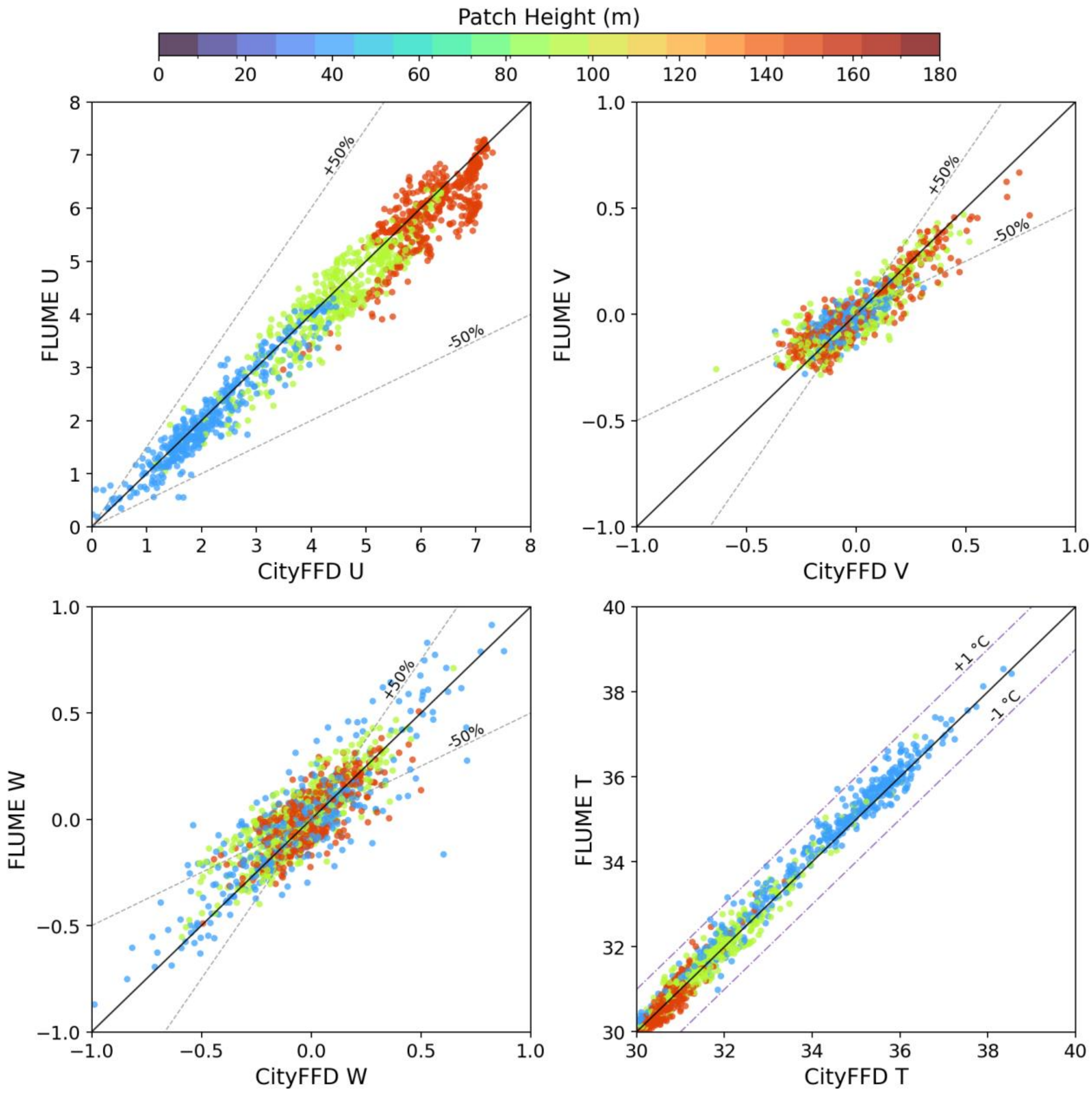

Figure 11: Correlation of U, V, W, T at patches from CityFFD and FLUME for Case 32

### 3.3. Uncertainty quantification framework

To assess the confidence of FLUME beyond the validation of Sections 3.1 and 3.2, an uncertainty quantification framework was constructed by combining deep ensembles with conformal prediction. Five FLUME models were trained using different partitions of training, validation, and testing cases. For each unseen case, the ensemble mean was used as the point prediction while the ensemble standard deviation served as a practical measure of sensitivity to the data split and as a proxy for epistemic uncertainty (Lakshminarayanan et al., 2017; Gawlikowski et al., 2023). Conformal prediction (Shafer and Vovk, 2007; Angelopoulos and Bates, 2021) was then applied by calibrating normalized residuals on the remaining cities in a leave-one-city-out manner, so that the final prediction intervals are directly related to the local spread of the ensemble. Two miscoverage levels, $\alpha = 0.30$ and $\alpha = 0.70$ (nominal coverage of 70% and 30%), were considered to examine both a wider and a narrower prediction interval.

The resulting intervals were evaluated using spatial maps, patch-level statistics, and vertical profiles at selected locations. In this way, the UQ framework does not only provide a global statement of reliability but also reveals where FLUME is most stable, where it is most sensitive to the choice of training subsets, and whether larger ensemble spread is actually associated with larger prediction error. The above is a fundamental requirement for any useful uncertainty estimate (Nguyen et al., 2021).

In Figure 12 the five combinations of training, validation, and testing cases used to construct the ensemble are presented. The unseen cases 0, 8, 20, 24 and 32 remain excluded from the corresponding learning stage, which allows the ensemble spread to reflect actual sensitivity to the selected subsets rather than information leakage. It is important to note that the combinations still share a large common portion of the dataset, thus the ensemble is not expected to generate completely different solutions, but rather realistic variations around a common learned aerodynamic behavior. This is useful, since if the combinations were too similar the ensemble spread would become artificially small, whereas if they were too different the spread would reflect inconsistency of the setup rather than epistemic uncertainty (Fort et al., 2019). Therefore, the five combinations define a balanced framework in which the ensemble members differ enough to reveal genuine sensitivity to the training data selection yet share enough common structure to ensure that the resulting spread reflects epistemic uncertainty rather than inconsistency of the learning setup.

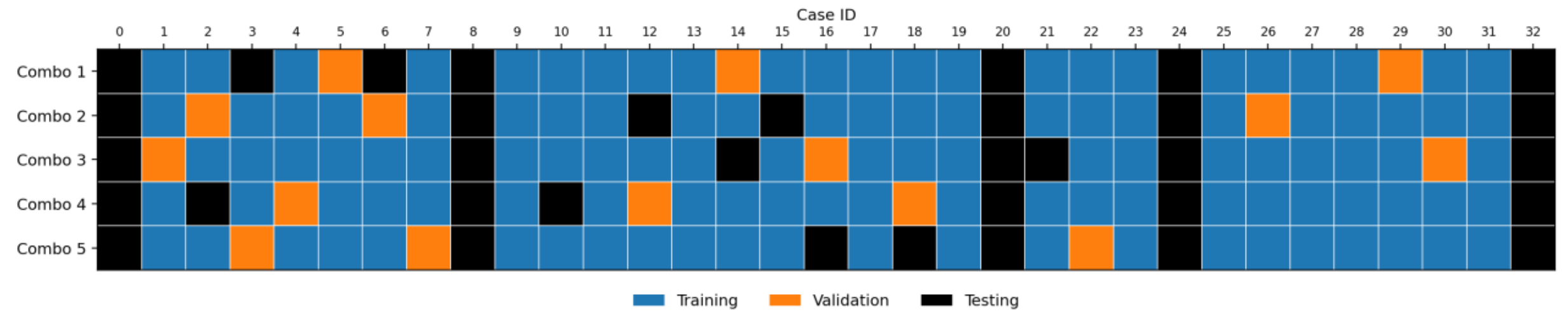

*Figure 12: Five training/validation/testing combinations of the 33 cases in the dataset used to construct the ensemble*

In Figure 13 the global metrics of the five ensemble members for all five unseen cases are presented and provide a first indication of the stability of FLUME. Regarding U and T, the curves across the five combinations remain relatively close to each other, particularly in terms of $R^2$, which suggests that the dominant streamwise flow organization and the temperature field are learned in a robust way and are not strongly dependent on the exact training subset. However, the spread between combinations becomes more visible in the RMSE and MAE results, especially for Cases 0 and 32 in U, which indicates that the local magnitude of errors can vary noticeably even when the global correlation structure is stable. Regarding V, differences between combinations are smaller in absolute terms, but this should be interpreted in the context of the overall lower accuracy of this component, rather than as evidence of stability. Regarding W, a similar behavior is observed, with moderate variation between combinations and overall lower accuracy than U and T. This is consistent with the higher sensitivity of vertical motions to local wake interactions, as also discussed in the previous sections. Regarding T, Case 20 shows a visible increase in RMSE and MAE across all combinations while $R^2$ remains very high. This suggests that FLUME preserves the thermal structure well but with a larger local magnitude error for that configuration. Clearly, the ensemble members provide consistent global conclusions while not being identical, which is a necessary condition for a meaningful ensemble-based uncertainty estimate.

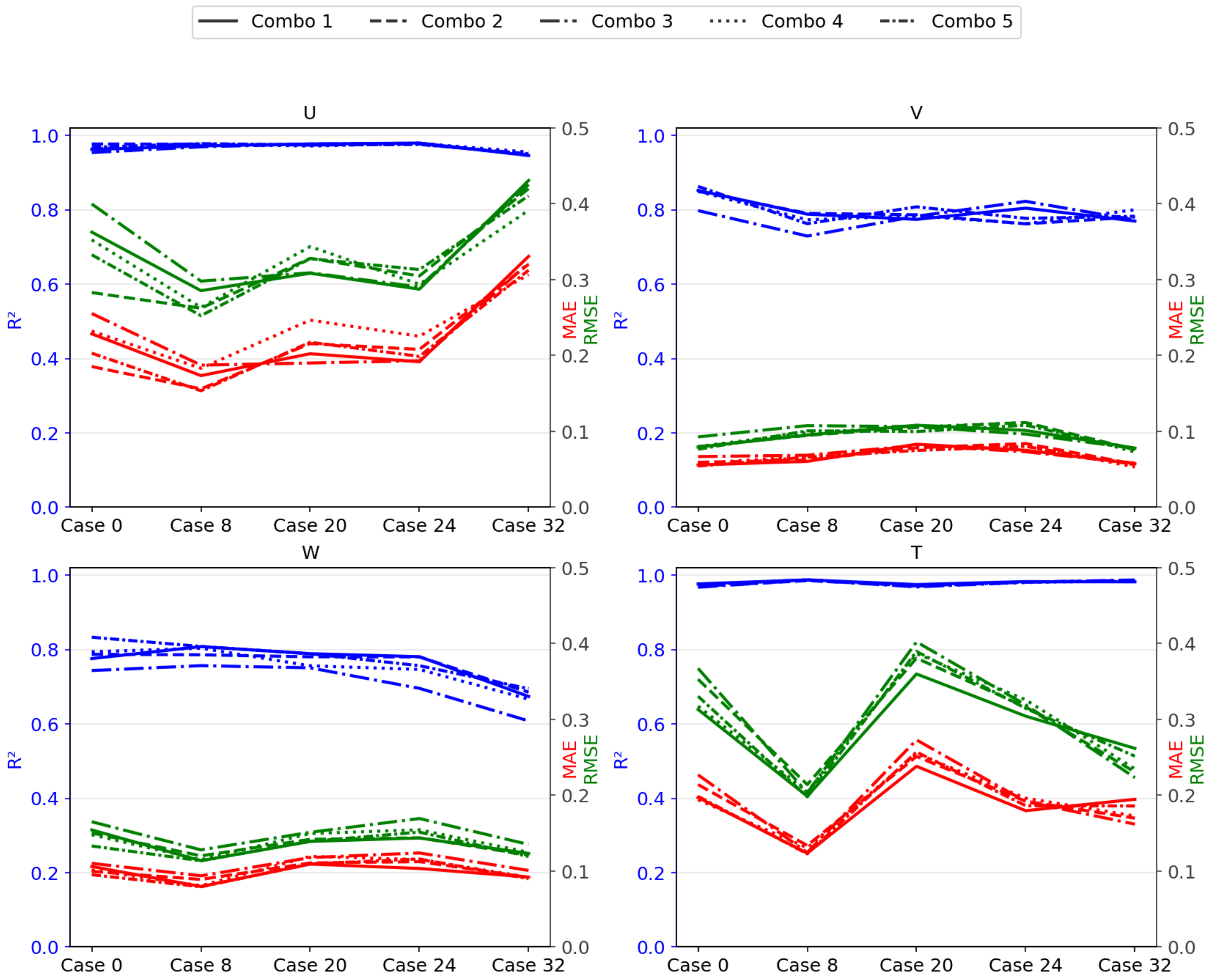

Figure 13: Accuracy metrics of five cases and five combinations

In Figure 14 the patch-mean ensemble standard deviation is compared with the patch-mean absolute error in log-log space. A clear positive relationship is observed in all panels, which indicates that the ensemble spread is informative of the local prediction error. This result is consistent with findings for neural operator ensembles in the broader scientific machine learning literature (Pensoneault and Zhu, 2025; Lakshminarayanan et al., 2017; Fort et al., 2019). The exponents of the fitted curves range from approximately 1.1 for W to 1.6 for U and T, indicating a slightly super-linear relationship. This means that patches where the ensemble is more uncertain tend to accumulate disproportionately larger errors, which is physically consistent with the fact that the most aerodynamically complex regions of the urban canopy simultaneously challenge both the mean prediction and the internal consistency of the model across training subsets.

Regarding U and V, the point clouds are broad but centered around the trend lines, and the coloring by patch height does not reveal a systematic pattern. Warm and cool colors, that represent large and small heights based on the colormap, are interleaved throughout the error-uncertainty space without a clear stratification. An additional search for correlation with patch occupancy, defined as the fraction of solid building volume within a patch, yielded similarly inconclusive results. This is not entirely surprising, since in realistic urban morphologies, the aerodynamic complexity experienced by horizontal velocity components is governed less by height alone and more by the local three-dimensional arrangement of buildings, their relative heights, spacing, and orientation with respect to the wind. For example, a patch at low height in an open street canyon and a patch at the same height immediately downstream of a tall, isolated building can occupy entirely different aerodynamic regime. Thus, neither height nor occupancy alone is sufficient to predict where errors will concentrate. The heterogeneity of the urban configurations in the dataset, which is precisely what makes FLUME generalization-relevant, also makes it difficult to isolate any single geometric descriptor as a reliable predictor of local prediction difficulty for the horizontal components.

Regarding W, a distinctly different pattern emerges. The lower-height patches, shown in darker purple and blue colors, tend to cluster toward the higher end of both the error and ensemble standard deviation axes, while patches at greater heights are more broadly distributed across the plot. This behavior is physically meaningful, since within and just above the urban canopy, the vertical velocity component is governed by strong impingement, separation, and recirculation effects that are highly sensitive to the exact local building configuration. These are precisely the regions where small differences in the training data partition, which drives the ensemble spread, can produce noticeably different local predictions of W, since the model must reconstruct sharp upward and downward motions from limited and geometrically complex training examples. At heights well above the canopy, where the flow is less directly disturbed by individual buildings, the vertical velocity is weaker in magnitude and smoother in spatial structure, and both error and uncertainty correspondingly decrease.

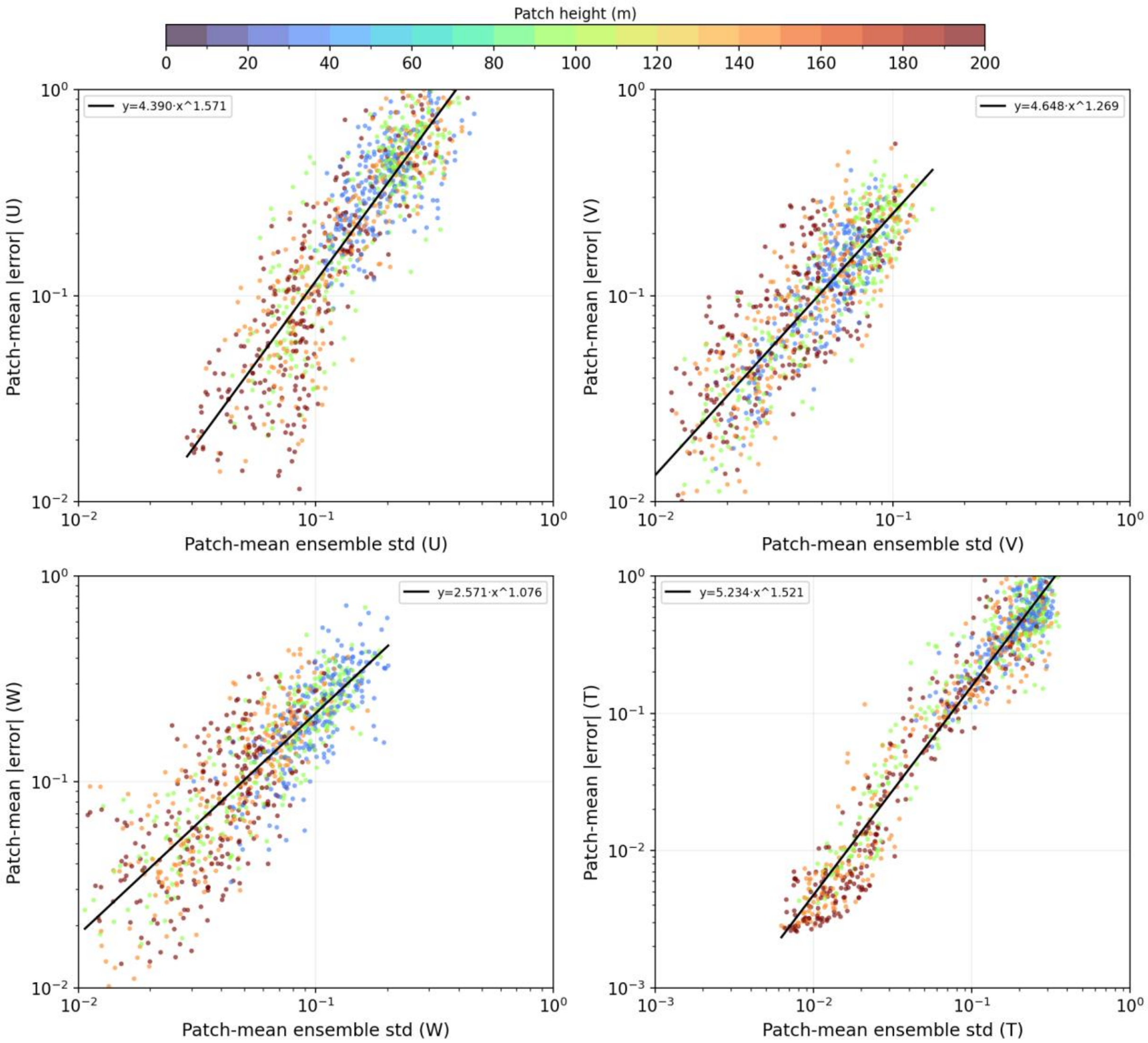

Figure 14: Patch-mean absolute error with respect to CityFFD versus patch-mean ensemble standard deviation

Regarding temperature T, the relationship is the tightest among all four variables. The coloring reveals a physically clear stratification, where the lowest errors concentrate at the highest patches while the largest errors are found near the urban surface. This is directly consistent with the surface-driven thermal forcing. Within the canopy the temperature field is strongly shaped by heated façades and local recirculation, whereas at greater heights it becomes progressively decoupled from individual building geometry as thermal plumes mix. This vertical gradient in thermal predictability is physically reasonable and provides additional confidence that the ensemble spread for T is tracking a genuine uncertainty signal rather than numerical noise.

In Figure 15 the vertical profiles of U, V, W, and T at the six locations A to F marked in the building cross-section at the top of the figure are presented, together with the CityFFD reference solution, the FLUME ensemble mean, and the conformal prediction intervals for $\alpha = 0.30$ and $\alpha = 0.70$. The six locations were selected to sample a representative range of aerodynamic environments: incoming flow close to buildings (A), locations near the wake of tall buildings and wind flow passages between buildings of various heights (B, C, F), and locations directly adjacent to the tallest structure in the domain (D, E).

The width of the confidence intervals follows a physically coherent pattern across all six locations. At A, where the flow approaches the urban cluster with limited geometric interference, the confidence intervals are the narrowest in the figure for all four variables. This indicates that FLUME is most confident in these aerodynamic regions, which is expected. As the profiles move into the urban canopy, the confidence changes substantially. At B, C, and F, that reside on building wakes and channeling passages, the intervals widen considerably for V and W, which reflects the greater sensitivity of these components to the exact building arrangement seen during training. The widening is maximized at C, where strong channeling between buildings produces complex flow features that the ensemble models resolve differently depending on the training subset. Location E lies in the immediate wake of the tallest building, where the intervals widen mainly near roof level, in the shear layer that separates from the building top. Location D lies within the footprint of the tallest building, so its profiles are only defined above the roof at about 170 m. At C and F, the CityFFD profiles fall outside the intervals over parts of the height for U, V, and W, which shows that the intervals do not cover all local errors in these wake and channeling regions. Regarding T, the intervals remain narrow at all six locations above rooftop level, with meaningful spread appearing only near the surface at B and C where heated façades and sheltering directly influence the local thermal field. Overall, these six profiles show that the width of the conformal prediction intervals follows the aerodynamic complexity of each location.

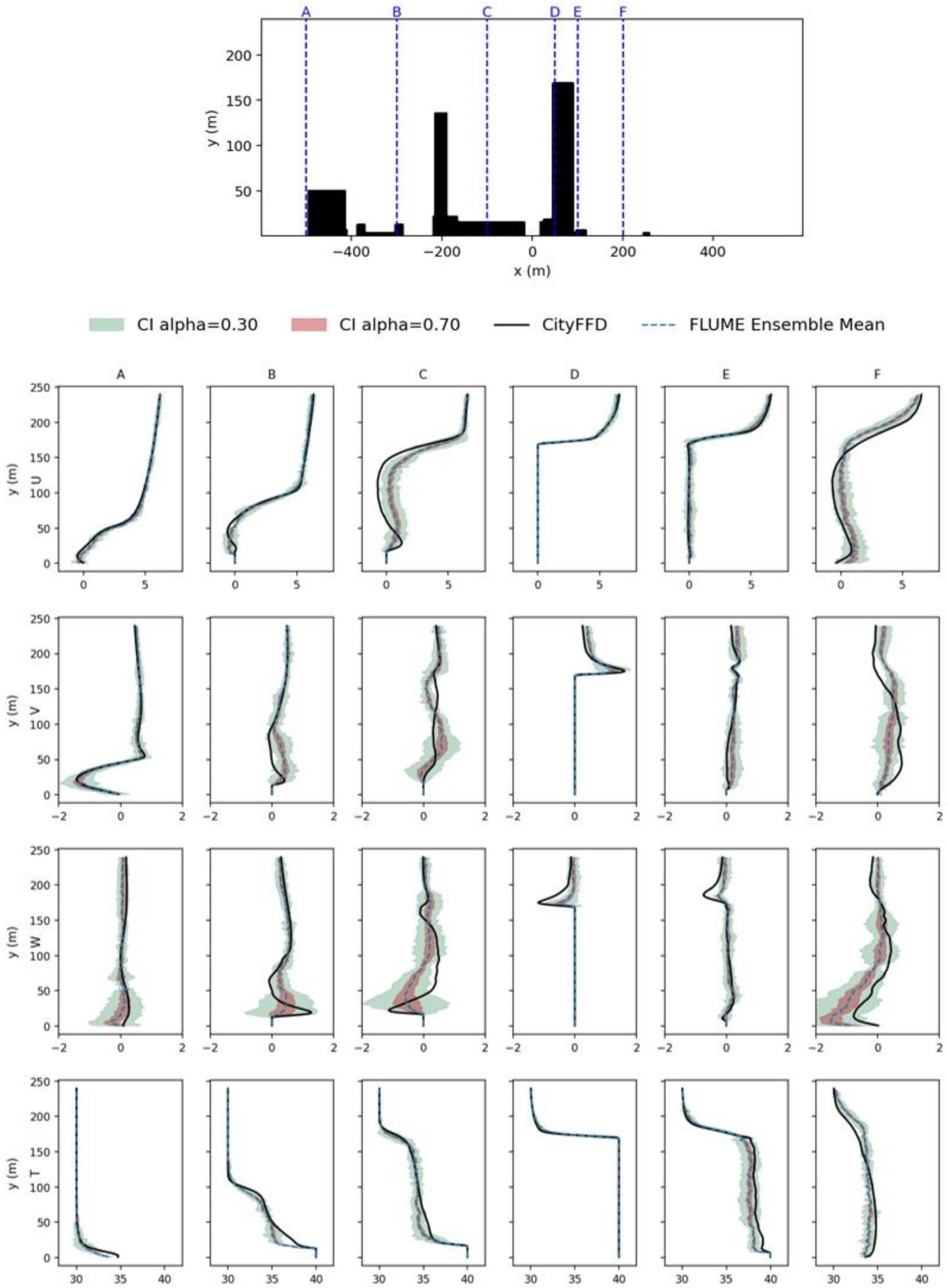

Figure 15: Vertical profiles with ensemble mean and conformal prediction intervals (α = 0.30 and 0.70) at selected locations

Across both miscoverage levels, the empirically observed coverage closely matches the nominal target (e.g. 66.6–75.8% at the 70% level and 28.8–32.3% at the 30% level for case 0), confirming that the conformal prediction framework is well calibrated. Notably, this close agreement is achieved despite calibrating on entirely separate cities (cases 8, 20, 24), which further supports that the ensemble standard deviation is an informative uncertainty proxy whose calibration transfers robustly across geometries. Together, these results support the practical reliability of the ensemble and conformal prediction UQ framework for FLUME.

## 4. Limitations and future considerations

FLUME, as presented in this paper, constitutes a first step toward a practical, geometry-aware, and uncertainty-aware fast urban microclimate emulator. The results presented in the previous sections are encouraging, although several limitations need to be acknowledged and addressed in future work to broaden its applicability and strengthen its reliability for engineering use.

The most direct limitation concerns the spatial resolution of the training data. The CityFFD simulations used here employ a near-building resolution of 4 m × 4 m × 1.5 m. Although this is sufficient to resolve the dominant flow patterns in the urban canopy, it does not capture the smaller-scale turbulent structures that are relevant for pedestrian-level comfort assessments in narrow streets or the detailed flow around individual façade elements. In the future, it would be beneficial to explore the coupling of FLUME with higher-resolution CFD data and examine how the predictive accuracy scales with grid refinement.

Another limitation is that the current framework has been trained and evaluated under fixed thermal forcing, and its applicability to different thermal boundary conditions and atmospheric stratification remains untested. Similarly, all simulations use a single inflow speed and direction, so the rotation and rescaling procedures described in Section 2.2.3 remain to be validated. Studies such as those of Tian et al. (2021), Marucci and Carpentieri (2020), and Uehara et al. (2000) have demonstrated the significant influence of thermal stratification on urban flow patterns, and this effect should eventually be represented in the FLUME framework. Besides that, the boundary conditions directly from WRF, similar to the work of Wang et al. (2023) and Mortezazadeh et al., (2021), can potentially be introduced as well.

The current approach does not parametrize the patch size or the overlap between adjacent patches. The patch dimensions (45 × 45 × 45) and the overlap strategy were selected based on preliminary testing but were not systematically optimized. A sensitivity study on these parameters, and potentially a learnable or adaptive patching strategy, could improve both accuracy in transition zones and computational efficiency. This should be investigated deeper in future work.

It is also important to note that the validation presented here relies exclusively on CityFFD as the reference solution, without direct comparison against physical wind tunnel or field measurements for realistic urban configurations. Although CityFFD has been validated against such measurements (Mortezazadeh et al., 2022; Mortezazadeh and Wang, 2020), the absence of a direct experimental benchmark for the specific configurations used in the validation of FLUME is a limitation. Wind tunnel campaigns on realistic urban configurations are needed for this purpose, covering various heights and not only the pedestrian level.

Regarding the UQ framework, the approach presented here is based on deep ensembles, which provide a practical but approximate proxy for epistemic uncertainty. Alternative approaches, including Monte Carlo dropout, Bayesian neural networks, and evidential deep learning, have been discussed in the literature (Nguyen et al., 2021; Gawlikowski et al., 2023) and could provide complementary or improved uncertainty estimates. Exploring these alternatives and comparing them systematically against the ensemble approach within the FLUME framework could be beneficial as to support more complete guidelines for UQ in fast urban microclimate predictions from machine learning.

Looking toward longer-term developments, several extensions of the current framework are of particular interest. The prediction of Reynolds stresses and turbulent kinetic energy would extend the framework beyond mean fields. As the field moves toward more standardized guidelines for the use of machine learning in urban microclimate prediction, the development of clear protocols for training data requirements, uncertainty reporting, and model validation should be of great importance for the community.

## 5. Conclusion

In this paper the Fast Localized Urban Microclimate Emulator (FLUME) was presented. FLUME is a rapid, data-efficient, and geometry-aware framework for the simultaneous prediction of three-dimensional mean wind velocity components and temperature in unseen urban morphologies. The framework combines FNO prediction on local 3D patches with the Multi-Directional Distance Feature (MDDF), a novel geometry encoding that captures directional distances to buildings across the full horizontal and vertical angular range. The model was trained on only 23 urban CFD cases generated by CityFFD. The main conclusions of the paper can be summarized as follows:

(a) The accuracy of FLUME was assessed against two unseen urban configurations, a 1.2 km × 1.2 km area of downtown Montreal and a larger 3 km × 3 km domain in Osaka. Mean absolute errors of 0.20 to 0.31 m/s for streamwise velocity and 0.18 to 0.19 °C for temperature were achieved, below the practical accuracy thresholds established in the wind engineering and urban microclimate literature.

(b) For the 1.2 km unseen configuration, $R^2$ reached 0.97 for both U and T, with FAC2 values of 0.98 and 1.00, respectively. Regarding the secondary velocity components, $R^2$ was 0.86 for V and 0.83 for W, with all MAE values remaining below the 0.5 m/s and 1 °C thresholds. For the 3 km unseen configuration, $R^2$ was 0.95 for U and 0.98 for T, with only a modest reduction in performance relative to the smaller domain. This indicates that the patch-based formulation allows FLUME to generalize to a larger urban domain, provided that the training data include at least one domain of comparable size.

(c) The hierarchy of predictive accuracy, highest for U and T, progressively lower for V and W, is physically consistent and has been observed in validation studies of LES and data-driven models for urban wind flows. Overall, FLUME achieves encouraging predictive accuracy given the limited training data.

(d) Regarding the uncertainty quantification, the framework based on deep ensembles combined with conformal prediction showed that the ensemble spread is positively correlated with the local prediction error across all variables and therefore serves as a meaningful proxy for epistemic uncertainty. The conformal prediction intervals were well calibrated, as shown by the close agreement between nominal and observed coverage, and adapted accurately to the local ensemble spread, providing spatially varying intervals that reveal where the FLUME predictions are least certain. These estimates followed the aerodynamic complexity of the flow, widening in regions such as building wakes and near shear layers, and thereby were used to quantify the level of confidence in the FLUME predictions.

(e) FLUME produces a full 3D prediction in 10 seconds after training, with total prediction time remaining under one minute when MDDF computation is included. This is a significant reduction compared to the computational cost of CFD, which is an encouraging outcome for the utilization of the method in practical applications.

Overall, the results presented in this paper indicate that FLUME is a promising tool for rapid urban microclimate prediction. The combination of strong generalization accuracy and an ensemble-based uncertainty framework makes it well suited for integration into urban design workflows, environmental wind engineering, and various applications where rapid prediction of wind flow and temperature patterns in urban environments is necessary to move towards a sustainable and energy efficient urban microclimate in cities.

# Acknowledgement

This research was supported by the Canada First Research Excellence Fund (Volt-Age) under the SEED project, "*Creating Electrified and Decarbonized Healthy Urban Microclimate around Building Clusters through Climate-Resilient Solutions*", and under the IMPACT project "*Transforming Built and Urban Microclimates: Advancing Resilience Science for Vulnerable Populations in a Decarbonized and Electrified Canada*". Numerical simulations were performed using Digital Research Alliance of Canada computing resources. We also acknowledge the review and comments from Dr. Di Zhou, Department of Mechanical and Aerospace Engineering, University of Tennessee, USA.